\documentstyle[12pt]{article}

\author{C. Bizdadea and S.O.Saliu\thanks{e-mail address:
odile@udjmath1.sfos.ro}\\
Department of Physics, University of Craiova\\
13 A.I.Cuza Str., Craiova R-1100, Romania}
\title{Extravariables in the BRST Quantization of Second-Class Constrained
Systems. Existence Theorems
}
\date{November 2, 1995
}
\def\cite#1{${\hbox{[#1]}}$}%
\input tcilatex
\newtheorem{theorem}{Theorem}

\begin{document}

\maketitle
\begin{abstract}
In this paper we show how the BRST quantization can be applied to systems
possessing only second-class constraints through their conversion to some
first-class ones starting with our method exposed in [Nucl.Phys. B456
(1995)473]. Thus, it is proved that i) for a certain class of second-class
systems there exists a standard coupling between the variables of the
original phase-space and some extravariables such that we can transform the
original system into a one-parameter family of first-class systems; ii) the
BRST quantization of this family in a standard gauge leads to the same path
integral as that of the original system. The analysis is accomplished in
both reducible and irreducible cases. In the same time, there is obtained
the Lagrangian action of the first-class family and its provenience is
clarified. In this context, the Wess-Zumino action is also derived. The
results from the theoretical part of the paper are exemplified in detail for
the massive Yang-Mills theory and for the massive abelian three-form gauge
fields.

PACS NUMBER: 11.10.Ef
\end{abstract}

\section{Introduction}

During the last years, the BRST method imposed itself as the only covariant
quantization method for gauge theories. It is well-known that at the
Hamiltonian level, to gauge theories correspond first (and eventually
second)-class constraints. The canonical quantization of the theories with
both first and second-class constraints has been accomplished in \cite{1}
and \cite{2}, while the BRST quantization of such theories is presented in
\cite{3}. A natural tendency is that of also quantizing the systems
possessing only second-class constraints in the BRST formalism. This cannot
be done directly because these theories do not possess gauge invariances.
This is why it is necessary to implement in the theory some gauge
invariances. This can be achieved by transforming the original second-class
system into a first-class one in the original phase-space \cite{4} or into a
larger one obtained from the original phase-space by introducing some extra
variables \cite{5}-\cite{6}. The BRST quantization of those second-class
systems whose constraint matrix does not depend on the canonical variables
is shown in \cite{4} and is realized through implementing some gauge
invariances in the original phase-space. Many authors \cite{7}-\cite{16}
have applied the methods from \cite{5}-\cite{6} and succeeded in quantizing
(in the BV, BRST or other methods) various models. The BRST quantization of
second-class systems in a larger phase-space has not been gained in a
general manner up to present. This is actually the purpose of our work.
Namely, in this paper it will be shown how to realize in general the BRST
quantization in a larger phase-space for systems subject only to
second-class constraints. More precisely, starting with an original
second-class system, we shall implement the following steps: i) we shall
transform this system into a first-class one in the original phase-space
\cite{4}; ii) from this last system we shall build a one-parameter family of
first-class systems in a larger phase-space in the case of irreducible
original second-class constraints, as well as in the case where these
initial constraints preserve somehow the trace of reducibility of a certain
first-class system; iii) we shall quantize the first-class family in the
light of the BRST formalism, obtaining in the end that its path integral is
identical with the one of the original system. This is the meaning of
applying the BRST quantization to second-class systems. We mention that our
method of turning the original second-class system into a first-class family
to be employed in step ii) is different from that exposed in \cite{5}-\cite
{6}. In this paper we use for the sake of simplicity the notations of
finite-dimensional analytical mechanics, but the analysis can be
straightforwardly extended to field theory. Related to the BRST
quantization, we follow the same lines as in \cite{17}.

The paper is organized into seven sections. In Sec.2 we shall briefly review
the BRST quantization of second-class constrained systems in the original
phase-space. Sec.3 is devoted to the construction of the one-parameter
family of first-class systems. There it will be proved the existence of the
Hamiltonian of the first-class family and it will be obtained its concrete
form. In Sec.4 we shall quantize in the antifield BRST formalism the
first-class family and prove that its path integral coincides with the one
of the original system. Sec.5 focuses on the Lagrangian approach of the
first-class family. Here it will be inferred the Lagrangian form of the path
integral for the first-class family under some simple assumptions and it
will be clarified the origin of this family. The Wess-Zumino action \cite{18}
associated with the introduction of extravariables is also emphasised. In
Sec.6 there will be exposed two examples illustrating the results derived in
the theoretical part of the paper. Sec.7 outlines some conclusions.

\section{The BRST quantization of second-class systems in the original
phase-space}

We follow the presentation of Ref. \cite{4}, to which we refer for details
and proofs. Our starting point is represented by a system with the canonical
Hamiltonian $H$, described by $N$ canonical pairs $(q^i,p_i)$, and subject
to the second-class constraints $\chi _\alpha =0$, where $\chi _\alpha
=\left( G_a,C_a\right) $ such that the constraint functions $G_a$ to satisfy
\begin{equation}
\label{(1)}\left[ G_a,G_b\right] =C_{ab}^{\quad c}G_c.
\end{equation}
The symbol $\left[ ,\right] $ denotes the Poisson bracket. Because the
constraint functions $\chi _\alpha $ are second-class, it results simply
from (\ref{(1)}) that
\begin{equation}
\label{(2)}\det C_{\alpha \beta }=\left( \det \left( \Delta _{ab}\right)
\right) ^2\neq 0,
\end{equation}
where $C_{\alpha \beta }=\left[ \chi _\alpha ,\chi _\beta \right] $ and $%
\Delta _{ab}=\left[ C_a,G_b\right] $. We treat only the case where $\Delta
_{ab}$'s do not depend on the canonical variables.

The first step in our quantization procedure consists in the construction of
a first-class Hamiltonian with respect to the functions $G_a$. Related to
this matter, the next theorem holds.

\begin{theorem}
Let $H$ be the canonical Hamiltonian of the system subject to the
second-class constraints, $\chi _\alpha =0$. Then, there exists a function $
\overline{H}=H+$``extraterms\ in\ $q$'s and $p$'s'' such that $\overline{H}$
is first-class with respect to the constraints $G_a=0$
\begin{equation}
\label{(3)}\left[ \overline{H},G_a\right] =f_{\;a}^bG_b,
\end{equation}
with $f_{\;a}^b$ some functions of $q$'s and $p$'s.
\end{theorem}

\TeXButton{Proof}{\proof} The proof is given in \cite{4}.$\Box $

The concrete form of $\overline{H}$ reads \cite{4}
\begin{eqnarray}\label{(4)}
\overline{H}&=&H+\sum\limits_{k=0}^\infty \frac{\left( -\right) ^{k+1}}{\left(
k+1\right) !}\left[ \ldots \left[ \left[ H,G_{m_{k+1}}\right] \Delta
^{m_{k+1}a_{k+1}},G_{m_k}\right] \Delta ^{m_ka_1},\ldots ,G_{m_1}\right]
\cdot\nonumber \\
&  &\Delta ^{m_1a_k}C_{a_1}C_{a_2}\ldots C_{a_{k+1}}+\lambda ^aG_a,
\end{eqnarray}
where $\Delta ^{ab}$ is the inverse of $\Delta _{ab}$, and $\lambda ^a$'s
are some functions taken such that $f_{\;a}^b=\lambda ^cC_{ca}^{\quad
b}+\left( -\right) ^{\epsilon _a\epsilon _b}\left[ \lambda ^b,G_a\right] $.
In the last formula, $\epsilon _a$ denotes the Grassmann parity of the
function $G_a$. Making a co-ordinate transformation of the type \cite{4}
\begin{equation}
\label{(5)}\left( q^i,p_i\right) \rightarrow \left( Q^a,P_a,z^\Delta ,
\overline{p}_\Delta \right) ,
\end{equation}
such that $P_a=G_a$, $Q^a=\Delta ^{ab}C_b$, $\left[ z^\Delta ,P_a\right]
=\left[ \overline{p}_\Delta ,P_a\right] =0,$ and $\left( z^\Delta ,\overline{%
p}_\Delta \right) $ to be canonical pairs, we associate to the original
system described by the action
\begin{equation}
\label{(6)}S_0\left[ q^i,p_i,\mu ^\alpha \right] =\int dt\left( \dot
q^ip_i-H-\mu ^\alpha \chi _\alpha \right) ,
\end{equation}
a first-class system with the action
\begin{equation}
\label{(7)}S_0\left[ Q^a,P_a,z^\Delta ,\overline{p}_\Delta ,v^a\right] =\int
dt\left( \dot Q^aP_a+\dot z^\Delta \overline{p}_\Delta -\tilde
H-v^aP_a\right) .
\end{equation}
In (\ref{(7)}), $\tilde H\left( Q^a,z^\Delta ,\overline{p}_\Delta \right) =
\overline{H}-\lambda ^aG_a=H\left( 0,z^\Delta ,\overline{p}_\Delta \right)
\equiv h\left( z^\Delta ,\overline{p}_\Delta \right) $, as deduced in \cite
{4}. Action (\ref{(7)}) is invariant under the gauge transformations $\delta
_\epsilon Q^a=\epsilon ^a$, $\delta _\epsilon v^a=\dot \epsilon ^a$, $\delta
_\epsilon z^\Delta =\delta _\epsilon \overline{p}_\Delta =\delta _\epsilon
P_a=0$. Transformation (\ref{(5)}) is not canonical in general, but its
Jacobian is equal to unity, as indicated in \cite{4}.

Let's pass now to the antifield BRST quantization of action (\ref{(7)}).
More precisely, we shall show that the path integral associated to action (
\ref{(6)}) is the same with the one corresponding to (\ref{(7)}) after its
BRST quantization in a gauge-fixing fermion implementing the canonical gauge
conditions $C_a=0$. These gauge conditions are equivalent to the conditions $%
Q^a=0$. The next theorem is helpful in finding the correct form of the above
mentioned gauge-fixing fermion.

\begin{theorem}
There exists a set of functions $f^a\left( Q\right) $ such that
\begin{equation}
\label{(8)}\frac 12\Delta _{ab}f^a\left( Q\right) f^b\left( Q\right)
=\sum\limits_{k=0}^\infty \frac{\left( -\right) ^{k+1}}{\left( k+1\right) !}
\frac{\overrightarrow{\partial }^{k+1}H}{\partial Q^{a_1}\ldots \partial
Q^{a_{k+1}}}Q^{a_1}\ldots Q^{a_{k+1}}.
\end{equation}
\end{theorem}

\TeXButton{Proof}{\proof} The proof is given in \cite{4}.$\Box $

The form of the functions $f^a\left( Q\right) $ reads \cite{4}
\begin{eqnarray}\label{(9)}
f^a\left( Q\right)&=&Q^a-\frac 1{3!}\Delta ^{(ab)}\left. \frac{
\overrightarrow{\partial }^3H}{\partial Q^b\partial Q^{b_1}\partial Q^{b_2}}
\right| _{Q=0}Q^{b_1}Q^{b_2}+\nonumber \\
&  &\sum\limits_{k=3}^\infty \frac 1{k!}\left. \frac{\overrightarrow{
\partial }^kf^c\left( Q\right) }{\partial Q^{c_1}\ldots \partial Q^{c_k}}
\right| _{Q=0}Q^{c_1}\ldots Q^{c_k},
\end{eqnarray}
where
\begin{eqnarray}
& &\frac 1{k!}\left. \frac{\overrightarrow{\partial }^kf^c\left( Q\right) }{
\partial Q^{c_1}\ldots \partial Q^{c_k}}\right| _{Q=0}=-\Delta ^{(cm)}\left(
\left. \frac 1{\left( k+1\right) !}\frac{\overrightarrow{\partial }^{k+1}H}{
\partial Q^m\partial Q^{c_1}\ldots \partial Q^{c_k}}\right| _{Q=0}+\right.
\nonumber \\
& &\left. \frac 1{2!}\Delta _{ab}\sum\limits_{j=2}^{k-1}\frac 1{j!\left(
k-j+1\right) !}\left. \frac{\overrightarrow{\partial }^jf^a\left( Q\right) }{
\partial Q^m\partial Q^{c_1}\ldots \partial Q^{c_{j-1}}}\right| _{Q=0}\left.
\frac{\overrightarrow{\partial }^{k-j+1}f^b\left( Q\right) }{\partial
Q^{c_j}\ldots \partial Q^{c_k}}\right| _{Q=0}\right) .\nonumber
\end{eqnarray}
In the last formulas, $\Delta ^{(ab)}$ is the inverse of the matrix $\Delta
_{(ab)}=\frac 12\left( \Delta _{ab}+\Delta _{ba}\right) $. It appears
clearly from (\ref{(9)}) that $f^a\left( Q\right) =0$ implies $Q^a=0$.
Taking the gauge-fixing fermion of the form $\Psi =-\int dt\left( \overline{%
\eta }^a\Delta _{ab}f^b\left( Q\right) \right) $, we obtain
\begin{equation}
\label{(10)}Z_\Psi =Z,
\end{equation}
where $Z_\Psi $ is the path integral of the first-class system in the gauge $%
\Psi $, and $Z$ is the path integral of the original system and is given by
\cite{17}
\begin{equation}
\label{(11)}Z=\int {\cal D}q\,{\cal D}p\,{\cal D}\mu \,\left( \det C_{\alpha
\beta }\right) ^{1/2}\exp \left( iS_0\left[ q^i,p_i,\mu ^\alpha \right]
\right) .
\end{equation}
At the level of independent variables $\left( z^\Delta ,\overline{p}_\Delta
\right) $, formula (\ref{(11)}) becomes
\begin{equation}
\label{indep}Z=\int {\cal D}z^\Delta \,{\cal D}\,\overline{p}_\Delta \exp
\left( i\int dt\left( \dot z^\Delta \overline{p}_\Delta -h\left( z^\Delta ,
\overline{p}_\Delta \right) \right) \right) .
\end{equation}

\section{The construction of the one-parameter family of first-class systems}

Within this section we shall extend the original phase-space and shall
construct in the new phase-space a one-parameter family of first-class
systems associated to the original theory. In the sequel we shall consider
only those purely bosonic systems with the primary constraints $G_a=0$ and
the secondary ones $C_a=0$, and whose phase-space is described by the real
functions $\left( q^i,p_i\right) $. The case of the systems having only
primary second-class constraints is treated in \cite{19}. We make, without
affecting the generality, the assumption that the functions $C_a$ can be
written under the form
\begin{equation}
\label{(12)}C_a=C_a^0+C_a^1,
\end{equation}
such that $\left[ C_a^0,C_b^0\right] =\left[ G_a,C_b^0\right] =0$ strongly,
and $\left[ C_a^1,G_b\right] =\Delta _{ab}$. Indeed, the form (\ref{(12)})
is not an additional restriction because if we make the transformation (\ref
{(5)}) we can always take $C_a^0=m_a\left( \overline{p}_\Delta \right) $ and
$C_a^1\equiv C_a-m_a\left( \overline{p}_\Delta \right) $, for some functions
$m_a$. In order to build the first-class family invoked above, the next
theorem is crucial in order to couple the original variables with the ones
to be added below.

\begin{theorem}
Let $H$ be the canonical Hamiltonian of a system possessing the primary
second-class constraints $P_a=0$, and the secondary ones $Q_a=0$. Then,

i) the sole real solution of the system $f^a\left( Q\right) =0$, where $%
f^a\left( Q\right) $ fulfill (\ref{(8)}), is $Q^a=0$;

ii) $\det \left( \frac{\partial f^a\left( Q\right) }{\partial Q^b}\right)
\neq 0$, for every $Q^a$ real.
\end{theorem}

\TeXButton{Proof}{\proof} i) From (\ref{(9)}) it results directly that $%
Q^a=0 $ is solution for $f^a\left( Q\right) =0$. It remains to prove that
this is the only real solution. Representing the canonical Hamiltonian as a
series of powers in $Q^a$'s, \cite{4}
\begin{eqnarray}\label{(13)}
 & &H\left( \left( Q^a,z^\Delta ,\overline{p}_\Delta \right) \right) =
H\left( 0,z^\Delta ,\overline{p}_\Delta \right) +\nonumber \\
 & &\sum\limits_{j=2}^\infty \frac 1{j!}\left. \frac{\partial ^jH\left(
Q^a,z^\Delta ,\overline{p}_\Delta \right) }{\partial Q^{a_1}\ldots
\partial Q^{a_j}}\right| _{Q=0}Q^{a_1}\ldots Q^{a_j},
\end{eqnarray} and introducing (\ref{(13)}) in (\ref{(8)}), we obtain
\begin{equation}
\label{(14)}\frac 12\Delta _{ab}f^a\left( Q\right) f^b\left( Q\right)
=H\left( 0,z^\Delta ,\overline{p}_\Delta \right) -H\left( Q^a,z^\Delta ,
\overline{p}_\Delta \right) .
\end{equation}
Differentiating (\ref{(14)}) with respect to $Q^c$, it follows
\begin{equation}
\label{(15)}\frac 12\Delta _{(ab)}\frac{\partial f^a\left( Q\right) }{%
\partial Q^c}f^b\left( Q\right) =-\frac{\partial H\left( Q^a,z^\Delta ,
\overline{p}_\Delta \right) }{\partial Q^c}.
\end{equation}
On the other hand, as $Q^a=0$ are a consequence of the constraints $P_a=0$,
it results
\begin{equation}
\label{(16)}\dot P_c=-\frac{\partial H\left( Q^a,z^\Delta ,\overline{p}%
_\Delta \right) }{\partial Q^c}=0\Longrightarrow Q^c=0.
\end{equation}
Comparing (\ref{(15)}) and (\ref{(16)}), it is clear that if there exists an
other real solution of $f^a\left( Q\right) =0$ than $Q^a=0$, e.g. $Q_0^a\neq
0$, then the system will also have the secondary constraints $Q_0^a=0$,
which contradicts the hypotheses. Thus, i) is proved.

ii) From (\ref{(9)}) we get $\left. \det \left( \frac{\partial f^a\left(
Q\right) }{\partial Q^b}\right) \right| _{Q=0}=1$, so the last determinant
is non-vanishing. In this way, it remains to be proved that
\begin{equation}
\label{(17)}\left. \det \left( \frac{\partial f^a\left( Q\right) }{\partial
Q^b}\right) \right| _{Q\neq 0}\neq 0,\qquad for\;every\;Q^a\;real.
\end{equation}
Using the result from i), it follows that we can represent $f^a\left(
Q\right) $ under the form
\begin{equation}
\label{(18)}f^a\left( Q\right) =V_{\quad b}^aQ^b,
\end{equation}
where $V_{\quad b}^a$ is an invertible matrix depending on $Q^a$'s, $%
z^\Delta $'s and $\overline{p}_\Delta $'s. With the aid of the fact that $%
Q^a=0$ are the only secondary constraints, we have
\begin{equation}
\label{(19)}-\frac{\partial H\left( Q^a,z^\Delta ,\overline{p}_\Delta
\right) }{\partial Q^c}=\overline{V}_{cb}Q^b,
\end{equation}
with $\overline{V}_{cb}$ an invertible matrix depending on the same
variables as $V_{\quad b}^a$. Introducing (\ref{(18)}) and (\ref{(19)}) in (
\ref{(15)}), we infer
\begin{equation}
\label{(20)}\frac 12\Delta _{(ab)}V_{\quad c}^b\frac{\partial f^c\left(
Q\right) }{\partial Q^d}=\overline{V}_{cd},\qquad for\;every\;real\;Q^a\neq
0.
\end{equation}
Taking the determinant in both hands of (\ref{(20)}) and taking into account
that the $V_{\quad c}^b$'s and $\overline{V}_{cd}$ are both invertible, it
results immediately (\ref{(17)}). This proves ii).$\Box $

It is easy to see that ii) implies $\det \left( \frac{\partial f^a\left(
C\right) }{\partial C_b}\right) \neq 0$, for every real functions $C_b$. In
the last relation, $f^a\left( C\right) $ is obtained from $f^a\left(
Q\right) $ using $Q^a=\Delta ^{ab}C_b$. The importance of the last theorem
resides in the fact that we can implement in a simple manner some secondary
first-class constraints $\gamma _a=0$ for the first-class family we intend
to construct through the term $\gamma _af^a\left( C\right) $ which we shall
introduce in the Hamiltonian of this family. It is precisely this term which
will couple the original variables with the new ones. This way of coupling
is one of the main points in our approach and reveals the main difference
between our conversion method and the one presented in \cite{5}-\cite{6}.
Indeed, choosing $\gamma _a$ such that $\left[ G_a,\gamma _b\right] =0$
strongly, we infer that $\left[ G_a,\gamma _bf^b\left( C\right) \right]
=-\Delta _{ca}\frac{\partial f^b\left( C\right) }{\partial C_c}\gamma _b$.
If we succeed in finding a Hamiltonian $H^{*}$ for the first-class family
satisfying $\left[ H^{*},G_a\right] =\left[ \gamma _bf^b\left( C\right)
,G_a\right] $, then the consistency of the primary constraints $G_a=0$ will
imply the secondary constraints $\gamma _a=0$ as $\det \left( \frac{\partial
f^b\left( C\right) }{\partial C_c}\right) \neq 0$. This problem will be
treated in the next two subsections. The splitting (\ref{(12)}) of the
functions $C_a$ will evidence two important cases, namely the irreducible
case where the functions $C_a^0$ are all independent, and the reducible case
where these functions are reducible. These cases will be treated separately.

\subsection{The irreducible case}

In this subsection we shall consider the case of the functions $C_a^0$ being
all independent. Then, the construction of the first-class family goes as
follows. For every pair $\left( G_a,C_a\right) $ we introduce a canonical
pair $\left( z^a,\overline{p}_a\right) $, so the new phase-space will have
the local co-ordinates $\left( q^i,p_i,z^a,\overline{p}_a\right) $. We
construct the Hamiltonian $H^{*}$ of the first-class family such that the
gauge algebra to be
\begin{equation}
\label{(21)}\left[ G_a,H^{*}\right] =-\Delta _{ca}\frac{\partial f^b\left(
C\right) }{\partial C_c}\gamma _b,
\end{equation}
\begin{equation}
\label{(22)}\left[ \gamma _a,H^{*}\right] =\left[ C_a^0,f^b\left( C\right)
\right] \gamma _b,
\end{equation}
\begin{equation}
\label{(23)}\left[ G_a,\gamma _b\right] =0,
\end{equation}
with the choice
\begin{equation}
\label{(24)}\gamma _a\equiv C_a^0+\lambda \overline{p}_a,
\end{equation}
$\lambda $ being the non-vanishing parameter of the first-class family. We
take the Hamiltonian $H^{*}$ of the form
\begin{equation}
\label{(25)}H^{*}=H^{\prime }-\frac{\lambda ^2}2\Delta ^{ab}\overline{p}_a
\overline{p}_b+\gamma _af^a\left( C\right) +g\left( q^i,p_i,z^a,\overline{p}%
_a\right) ,
\end{equation}
where
\begin{equation}
\label{(26)}H^{\prime }=\tilde H+\frac 12\Delta ^{ab}C_a^0C_b^0,
\end{equation}
and $g\left( q^i,p_i,z^a,\overline{p}_a\right) $ is a function to be further
derived. We notice that the first piece of $H^{*}$ is the Hamiltonian of the
first-class system in the original phase-space associated to the original
theory. In fact, $H^{\prime }$ and $\tilde H$ are in the same class of gauge
invariant functions with respect to the $G_a$'s. In principle, the second
term in the right hand of (\ref{(26)}) can be taken any function of $C_a^0$%
's. As it will be seen, the necessity of taking this term quadratic in $%
C_a^0 $'s is directly connected with the choice of the quadratic term in the
momenta $\overline{p}_a$'s from (\ref{(25)}). This last quadratic term is
motivated by a simpler passing to the Lagrangian formalism (see Sec.5).
Replacing ( \ref{(25)}) in (\ref{(21)}) and (\ref{(22)}), we get the
equations
\begin{equation}
\label{(27)}\left[ G_a,g\left( q^i,p_i,z^a,\overline{p}_a\right) \right] =0,
\end{equation}
\begin{equation}
\label{(28)}\left[ C_a^0,H^{\prime }\right] +\lambda \left[ \overline{p}%
_a,g\left( q^i,p_i,z^a,\overline{p}_a\right) \right] =0.
\end{equation}
Now, it appears more obviously the reason of choosing the gauge algebra of
the form (\ref{(21)}-\ref{(23)}), the secondary first-class constraints as
in ( \ref{(24)}) and the Hamiltonian $H^{*}$ like in (\ref{(25)}). With
these choices, the first-class family is determined from the original system
up to the function $g\left( q^i,p_i,z^a,\overline{p}_a\right) $, which has
to fulfill (\ref{(27)}-\ref{(28)}). The next theorem shows that this
function can be also completely gained from the original system.

\begin{theorem}
There exists a function $g\left( q^i,p_i,z^a,\overline{p}_a\right) $
satisfying (\ref{(27)}-\ref{(28)}).
\end{theorem}

\TeXButton{Proof}{\proof} The proof is intended to be constructive, finally
obtaining the concrete form of $g\left( q^i,p_i,z^a,\overline{p}_a\right) $.
We represent this function as a series of powers in $z^a$'s with
coefficients depending on $\left( q^i,p_i,\overline{p}_a\right) $
\begin{equation}
\label{(29)}g\left( q^i,p_i,z^a,\overline{p}_a\right)
=\sum\limits_{k=1}^\infty \stackrel{(k)}{g}_{a_1\ldots a_k}\left( q^i,p_i,
\overline{p}_a\right) z^{a_1}\ldots z^{a_k}.
\end{equation}
Inserting (\ref{(29)}) in (\ref{(28)}) and identifying the coefficients of
the same powers in $z^a$'s, we find the following tower of equations
\begin{equation}
\label{(30)}\lambda \stackrel{(1)}{g}_a=\left[ C_a^0,H^{\prime }\right] ,
\end{equation}
\begin{equation}
\label{(31)}2\lambda \stackrel{(2)}{g}_{a_1a_2}=\left[ C_{a_1}^0,\stackrel{%
(1)}{g}_{a_2}\right] ,
\end{equation}
$$
\vdots
$$
\begin{equation}
\label{(32)}k\lambda \stackrel{(k)}{g}_{a_1\ldots a_k}=\left[ C_{a_1}^0,%
\stackrel{(k-1)}{g}_{a_2\ldots a_k}\right] ,
\end{equation}
$$
\vdots
$$
Using (\ref{(30)}-\ref{(32)}), we deduce in a simple manner
\begin{equation}
\label{(33)}\stackrel{(k)}{g}_{a_1\ldots a_k}=\frac 1{k!\lambda ^k}\left[
C_{a_1}^0,\left[ C_{a_2}^0,\ldots ,\left[ C_{a_k}^0,H^{\prime }\right]
\ldots \right] \right] .
\end{equation}
In this way, we proved that (\ref{(29)}) with the coefficients (\ref{(33)})
is the solution of (\ref{(28)}). Using the Jacobi identity we get
immediately
\begin{equation}
\label{(34)}\left[ G_a,\stackrel{(k)}{g}_{a_1\ldots a_k}\right] =0,\qquad
for\;every\;k,
\end{equation}
so (\ref{(27)}) is also fulfilled. This ends the proof.$\Box $

Because $\left[ C_a^0,H^{\prime }\right] =\left[ C_a^0,\tilde H\right] $, we
finally obtain
\begin{equation}
\label{(35)}g\left( q^i,p_i,z^a,\overline{p}_a\right)
=\sum\limits_{k=1}^\infty \frac 1{k!\lambda ^k}\left[ C_{a_1}^0,\left[
C_{a_2}^0,\ldots ,\left[ C_{a_k}^0,\tilde H\right] \ldots \right] \right]
z^{a_1}\ldots z^{a_k}.
\end{equation}
In this way, we associated to the original system depicted by action (\ref
{(6)}) a one-parameter family of first-class systems described by the action
\begin{equation}
\label{(36)}S_0\left[ q^i,p_i,z^a,\overline{p}_a,v^a,u^a\right] =\int
dt\left( \dot q^ip_i+\dot z^ap_a-H^{*}-v^aG_a-u^a\gamma _a\right) .
\end{equation}
Action (\ref{(36)}) is invariant under the gauge transformations:
\begin{equation}
\label{(37)}\delta _\epsilon q^i=\left[ q^i,G_a\right] \epsilon _1^a+\left[
q^i,C_a^0\right] \epsilon _2^a,
\end{equation}
\begin{equation}
\label{(38)}\delta _\epsilon p_i=\left[ p_i,G_a\right] \epsilon _1^a+\left[
p_i,C_a^0\right] \epsilon _2^a,
\end{equation}
\begin{equation}
\label{(39)}\delta _\epsilon z^a=\lambda \epsilon _2^a,
\end{equation}
\begin{equation}
\label{(40)}\delta _\epsilon \overline{p}_a=0,
\end{equation}
\begin{equation}
\label{(41)}\delta _\epsilon v^a=\dot \epsilon _1^a-C_{bc}^{\quad
a}v^b\epsilon _1^c,
\end{equation}
\begin{equation}
\label{(42)}\delta _\epsilon u^a=\dot \epsilon _2^a+\left[ C_b^0,f^a\left(
C\right) \right] \epsilon _2^b-\Delta _{cb}\frac{\partial f^a\left( C\right)
}{\partial C_c}\epsilon _1^b.
\end{equation}
These gauge transformations will be used in Sec.4 to the BRST quantization
of the irreducible first-class family.

\subsection{The reducible case}

Within this subsection, we shall examine the case where the functions $C_a^0$
are not all independent. This means that there exist some functions on $q^i$%
's and $p_i$'s denoted by $Z_{\;a_1}^a$, not all vanishing, such that
\begin{equation}
\label{(43)}Z_{\;a_1}^aC_a^0=0.
\end{equation}
We note that relations (\ref{(43)}) represent some identities holding for
all $q^i$ 's and $p_i$'s. Taking the Poisson brackets of $G_a$'s and $C_a^0$%
's with both hands of (\ref{(43)}), we get the following identities
\begin{equation}
\label{(44)}\left[ G_b,Z_{\;a_1}^a\right] =0,
\end{equation}
\begin{equation}
\label{(45)}\left[ C_b^0,Z_{\;a_1}^a\right] =0.
\end{equation}
With the aid of (\ref{(45)}) we deduce, supposing that (\ref{(43)}) are the
sole reducibility relations for $C_a^0$'s, the identities
\begin{equation}
\label{(46)}\left[ Z_{\;a_1}^a,Z_{\;b_1}^b\right] =0.
\end{equation}
At this point we are able to build consistently the first-class family in
the reducible case. This construction goes as follows. For every pair $%
\left( G_a,C_a\right) $ we introduce a new canonical pair $\left( z^a,
\overline{p}_a\right) $ such that the consistency of the $G_a$'s to imply
the secondary constraints $\gamma _a=0$, with $\gamma ^a$ given by (\ref
{(24)}). From (\ref{(43)}) it results simply
\begin{equation}
\label{(47)}Z_{\;a_1}^a\gamma _a=\lambda Z_{\;a_1}^a\overline{p}_a=0.
\end{equation}
For every relation (\ref{(47)}) we add a new canonical pair $\left(
y^{a_1},\pi _{a_1}\right) $ together with the constraint
\begin{equation}
\label{(48)}\gamma _{a_1}\equiv \pi _{a_1}=0,
\end{equation}
such that the consistency of the last constraints to imply the secondary
ones of the form
\begin{equation}
\label{(49)}\overline{\gamma }_{a_1}\equiv -Z_{\;a_1}^a\overline{p}_a=0,
\end{equation}
which are precisely (\ref{(47)}) up to a factor. Through this mechanism we
cannot generate new constraints even if the reducibility functions $%
Z_{\;a_1}^a$ are not all independent. In general, we can assume that there
exist some non-vanishing functions of $\left( q^i,p_i\right) $, denoted by $%
Z_{\;_{a_2}}^{a_1},\ldots ,Z_{\;a_k}^{a_{k-1}}$, such that the next
identities to hold
\begin{equation}
\label{(50)}Z_{\;a_1}^aZ_{\;_{a_2}}^{a_1}=0,
\end{equation}
$$
\vdots
$$
\begin{equation}
\label{(51)}Z_{\;a_{k-1}}^{a_{k-2}}Z_{\;a_k}^{a_{k-1}}=0.
\end{equation}
{}From (\ref{(50)}) we draw that
\begin{equation}
\label{(52)}Z_{\;a_2}^{a_1}\overline{\gamma }_{a_1}=0.
\end{equation}
If we repeated now the procedure between formulas (\ref{(47)}-\ref{(49)}),
we would introduce some new canonical pairs $\left( x^{a_2},\Pi
_{a_2}\right) $ together with the constraints $\Pi _{a_2}=0$, such that
their consistency to induce the ``tertiary constraints''
\begin{equation}
\label{(53)}\tilde \gamma _{a_2}\equiv Z_{\;_{a_2}}^{a_1}\overline{\gamma }%
_{a_1}=0.
\end{equation}
Relations (\ref{(53)}) are not constraints but identities due to (\ref{(50)}%
). Thus, the maximal set of constraints we can generate through the above
procedure is given by $G_a=0$, $\gamma _a=0$ and (\ref{(48)}-\ref{(49)}).
{}From (\ref{(44)}-\ref{(46)}) it follows that all the previous constraints
are first-class. A new feature of these constraints is that they become
reducible. The reducibility relations read
\begin{equation}
\label{(54)}Z_{\;a_1}^a\gamma _a+\lambda \overline{\gamma }_{a_1}=0.
\end{equation}

In the sequel, we shall build the Hamiltonian $\tilde H^{*}$ of the
reducible first-class family to satisfy
\begin{equation}
\label{(55)}\left[ G_a,\tilde H^{*}\right] =-\Delta _{ca}\frac{\partial
f^b\left( C\right) }{\partial C_c}\gamma _b,
\end{equation}
\begin{equation}
\label{(56)}\left[ \gamma _{a_1},\tilde H^{*}\right] =\overline{\gamma }%
_{a_1},
\end{equation}
\begin{equation}
\label{(57)}\left[ \gamma _a,\tilde H^{*}\right] =\left[ C_a^0,f^b\left(
C\right) \right] \gamma _b,
\end{equation}
\begin{equation}
\label{(58)}\left[ \overline{\gamma }_{a_1},\tilde H^{*}\right]
=M_{\;a_1}^a\gamma _a,
\end{equation}
with $M_{\;a_1}^a$ some functions to be subsequently determined. In the
reducible case, we take the Hamiltonian $\tilde H^{*}$ of the form
\begin{equation}
\label{(59)}\tilde H^{*}=H^{\prime }-\frac{\lambda ^2}2\Delta ^{ab}\overline{%
p}_a\overline{p}_b+\gamma _af^a\left( C\right) -y^{a_1}\overline{\gamma }%
_{a_1}+\tilde g\left( q^i,p_i,z^a,\overline{p}_a,y^{a_1},\pi _{a_1}\right) ,
\end{equation}
with $\tilde g\left( q^i,p_i,z^a,\overline{p}_a,y^{a_1},\pi _{a_1}\right) $
a function to be further obtained. Inserting (\ref{(59)}) in (\ref{(55)}-\ref
{(57)}) we achieve the following equations
\begin{equation}
\label{(60)}\left[ G_a,\tilde g\left( q^i,p_i,z^a,\overline{p}_a,y^{a_1},\pi
_{a_1}\right) \right] =0,
\end{equation}
\begin{equation}
\label{(61)}\left[ \pi _{a_1},\tilde g\left( q^i,p_i,z^a,\overline{p}%
_a,y^{a_1},\pi _{a_1}\right) \right] =0,
\end{equation}
\begin{equation}
\label{(62)}\left[ C_a^0,H^{\prime }\right] +\lambda \left[ \overline{p}%
_a,\tilde g\left( q^i,p_i,z^a,\overline{p}_a,y^{a_1},\pi _{a_1}\right)
\right] =0.
\end{equation}
It is simply to observe that the function $g\left( q^i,p_i,z^a,\overline{p}%
_a\right) $ given by (\ref{(35)}) verifies automatically (\ref{(60)}-\ref
{(62)}). Thus, it is left to be shown that $g\left( q^i,p_i,z^a,\overline{p}%
_a\right) $ verifies also (\ref{(58)}). This is the aim of the next theorem.

\begin{theorem}
Let $f$ be a solution of equations (\ref{(60)}-\ref{(62)}). Then, $f$
satisfies (\ref{(58)}), where $\tilde H^{*}$ is given by (\ref{(59)}) with $%
\tilde g=f$.
\end{theorem}

\TeXButton{Proof}{\proof} Taking the Poisson bracket of both hands in (\ref
{(54)}) with $\tilde H^{*}$ we get
\begin{equation}
\label{(63)}\left[ Z_{\;a_1}^a\gamma _a,\tilde H^{*}\right] +\lambda \left[
\overline{\gamma }_{a_1},\tilde H^{*}\right] =0.
\end{equation}
{}From (\ref{(57)}) (which is verified if $f$ is solution for (\ref{(60)}-\ref
{(62)})) and (\ref{(63)}) we infer
\begin{equation}
\label{(64)}\left[ \overline{\gamma }_{a_1},\tilde H^{*}\right] =-\frac
1\lambda \left( Z_{\;a_1}^b\left[ C_b^0,f^a\left( C\right) \right] +\left[
Z_{\;a_1}^a,\tilde H^{*}\right] \right) \gamma _a.
\end{equation}
The last relations are nothing but (\ref{(58)}), with
\begin{equation}
\label{(65)}M_{\;a_1}^a=-\frac 1\lambda \left( Z_{\;a_1}^b\left[
C_b^0,f^a\left( C\right) \right] +\left[ Z_{\;a_1}^a,\tilde H^{*}\right]
\right) .
\end{equation}
This completes the proof.$\Box $

Now it is clear the reason for choosing $\tilde H^{*}$ to fulfill (\ref{(55)}%
-\ref{(58)}). Indeed, accordingly the above theorem equations (\ref{(55)}-%
\ref{(58)}) are compatible with
\begin{equation}
\label{(66)}\tilde g\left( q^i,p_i,z^a,\overline{p}_a,y^{a_1},\pi
_{a_1}\right) =g\left( q^i,p_i,z^a,\overline{p}_a\right) ,
\end{equation}
with $g$ given by (\ref{(35)}).

To conclude with, we associated to the original theory (\ref{(6)}) a
reducible first-class family described by the action%
\begin{eqnarray}
\label{(67)}
& &S_0^{\prime }\left[ q^i,p_i,z^a,\overline{p}_a,y^{a_1},\pi
_{a_1},v^a,v^{a_1},u^a,u^{a_1}\right] =\int dt\left( \dot q^ip_i+\right.
\nonumber \\
& &\left. \dot z^a\overline{p}_a+\dot y^{a_1}\pi _{a_1}-\tilde
H^{*}-v^aG_a-v^{a_1}\gamma _{a_1}-u^a\gamma _a-u^{a_1}\overline{\gamma }%
_{a_1}\right) .
\end{eqnarray}
The gauge invariances of action (\ref{(67)}) are deduced to be
\begin{equation}
\label{(68)}\delta _\epsilon q^i=\left[ q^i,G_a\right] \epsilon _1^a+\left[
q^i,C_a^0\right] \epsilon _2^a-\left[ q^i,Z_{\;a_1}^a\right] \overline{p}%
_a\epsilon _4^{a_1},
\end{equation}
\begin{equation}
\label{(69)}\delta _\epsilon p_i=\left[ p_i,G_a\right] \epsilon _1^a+\left[
p_i,C_a^0\right] \epsilon _2^a-\left[ p_i,Z_{\;a_1}^a\right] \overline{p}%
_a\epsilon _4^{a_1},
\end{equation}
\begin{equation}
\label{(70)}\delta _\epsilon z^a=\lambda \epsilon _2^a-Z_{\;a_1}^a\epsilon
_4^{a_1},
\end{equation}
\begin{equation}
\label{(71)}\delta _\epsilon \overline{p}_a=0,
\end{equation}
\begin{equation}
\label{(72)}\delta _\epsilon y^{a_1}=\epsilon _3^{a_1},
\end{equation}
\begin{equation}
\label{(73)}\delta _\epsilon \pi _{a_1}=0,
\end{equation}
\begin{equation}
\label{(74)}\delta _\epsilon v^a=\dot \epsilon _1^a-C_{bc}^{\quad
a}v^b\epsilon _1^c,
\end{equation}
\begin{equation}
\label{(75)}\delta _\epsilon v^{a_1}=\dot \epsilon _3^a,
\end{equation}
\begin{eqnarray}
\label{(76)}
& &\delta _\epsilon u^a=\dot \epsilon _2^a+\left[ C_b^0,f^a\left( C\right)
\right] \epsilon _2^b-\Delta _{cb}\frac{\partial f^a\left( C\right) }{
\partial C_c}\epsilon _1^b-\nonumber \\
& &\frac 1\lambda \left( Z_{\;a_1}^b\left[ C_b^0,f^a\left( C\right)
\right] +\left[ Z_{\;a_1}^a,\tilde H^{*}\right] \right) \epsilon
_4^{a_1}+Z_{\;a_1}^a\epsilon _5^{a_1},
\end{eqnarray}
\begin{equation}
\label{(77)}\delta _\epsilon u^{a_1}=\epsilon _3^{a_1}+\dot \epsilon
_4^a+\lambda \epsilon _5^{a_1}.
\end{equation}
The gauge parameters $\epsilon _5^{a_1}$ appear due to the reducibility
relations (\ref{(54)}) which allow us to introduce the additional gauge
invariances \cite{17}, \cite{20}
\begin{equation}
\label{(78)}\delta _\epsilon u^A=Z_{\;b_1}^A\epsilon _5^{b_1},
\end{equation}
where $u^A=\left( u^a,u^{a_1}\right) $, and $Z_{\;b_1}^A=\left(
Z_{\;b_1}^a,\lambda \delta _{\;b_1}^{a_1}\right) $ are the reducibility
functions from (\ref{(54)}). The gauge invariances (\ref{(68)}-\ref{(77)})
will be employed within the BRST quantization of the reducible first-class
family.

\section{The antifield BRST quantization of the first-class family}

In this section we shall quantize the first-class families constructed
earlier in the context of the antifield BRST formalism based on path
integrals. As there appear major differences between the reducible and
irreducible cases, we shall treat them separately.

\subsection{The quantization in the irreducible case}

The starting point is given by action (\ref{(36)}) together with the gauge
transformations (\ref{(37)}-\ref{(42)}). Because the constraints are
irreducible, the minimal ghost spectrum \cite{17} will contain only the
ghosts $\left( \eta _1^a,\eta _2^a\right) $ correspondent to the gauge
parameters $\left( \epsilon _1^a,\epsilon _2^a\right) $. The Grassmann
parities and ghost numbers of the above ghosts are all equal to one. For all
the variables $\Phi ^I=\left( q^i,p_i,z^a,\overline{p}_a,v^a,u^a\right) $ we
introduce the antifields \cite{17}
$$
\Phi _I^{*}=\left( q_i^{*},p^{*i},z_a^{*},\overline{p}^{*a},v_a^{*},u_a^{*}%
\right) ,
$$
all of Grassmann parity one and ghost number minus one. The non-minimal
sector is taken to contain the variables
$$
\left( B_1^a,B_{1a}^{*},B_2^a,B_{2a}^{*},\overline{\eta }_{1a}^{*},\overline{%
\eta }_1^a,\overline{\eta }_{2a}^{*},\overline{\eta }_2^a\right) .
$$
Then, the non-minimal solution of the master equation reads%
\begin{eqnarray}
\label{(79)}
& &S=S_0\left[ q^i,p_i,z^a,\overline{p}_a,v^a,u^a\right] +\int dt\left(
q_i^{*}\left( \frac{\partial G_a}{\partial p_i}\eta _1^a+\frac{
\partial C_a^0}{\partial p_i}\eta _2^a\right) -\right. \nonumber \\
& &p^{*i}\left( \frac{\partial G_a}{\partial q^i}\eta _1^a+\frac{
\partial C_a^0}{\partial q^i}\eta _2^a\right) +\lambda z_a^{*}
\eta _2^a+v_a^{*}\left( \dot \eta _1^a-C_{bc}^{\quad a}v^b\eta _1^c\right) +
\nonumber \\
& &u_a^{*}\left( \dot \eta _2^a+\left[ C_b^0,f^a\left( C\right) \right]
\eta _2^b-\Delta _{cb}\frac{\partial f^a\left( C\right) }{\partial C_c}
\eta _1^b\right) + \nonumber \\
& &\left. \overline{\eta }_{1a}^{*}B_1^a+\overline{\eta }_{2a}^{*}B_2^a+
\ldots \right) ,
\end{eqnarray} where ``$\ldots $'' signify the terms of antighost numbers
greater than one. These terms are not essential because of the special form
of the gauge-fixing fermion to be outlined below. The standard gauge-fixing
fermion in our methods reads
\begin{equation}
\label{(80)}\Psi ^{\prime }=-\int dt\left( \overline{\eta }_1^a\Delta
_{ab}f^b\left( \rho _c\right) -\lambda \overline{\eta }_2^a\Delta
_{ab}z^b\right) ,
\end{equation}
where
\begin{equation}
\label{(81)}\rho _c=C_c-\frac 1\lambda \left[ C_c,C_b^0\right] z^b.
\end{equation}
We observe that $\Psi ^{\prime }$ reduces to $\Psi $ (given in Sec.2) in the
absence of the extravariables ($z^a=0$). The gauge-fixing fermion (\ref{(80)}%
) implements the canonical gauge conditions $C_a=0$ and $z^a=0$. Eliminating
in the usual manner the antifields from (\ref{(79)}), we derive the next
gauge-fixed action%
\begin{eqnarray}\label{(82)}
S_{\Psi ^{^{\prime }}}&=&S_0\left[ q^i,p_i,z^a,\overline{p}_a,v^a,u^a\right] +
\int dt\left( \lambda ^2\overline{\eta }_2^a\Delta _{ab}\eta _2^b+\lambda
\Delta _{ab}B_2^az^b-\right. \nonumber \\
& &\overline{\eta }_1^a\Delta _{ab}\frac{\partial f^b\left( \rho \right) }{
\partial \rho _c}\left( \Delta _{cd}\eta _1^d-\frac 1\lambda z^d\left[ \left[
C_c,C_d^0\right] ,C_e^0\right] \eta _2^e\right) -\nonumber \\
& &\left. \Delta _{ab}B_1^af^b\left( \rho \right) \right) .
\end{eqnarray} Employing repeatedly the Jacobi identity together with the
fact that the term $\frac{\partial ^2f^a\left( \rho \right) }{\partial \rho
_b\partial \rho _c}$ is symmetric in $b$ and $c$, it is simply to see that (%
\ref{(82)}) is invariant under the following BRST transformations
\begin{equation}
\label{(83)}sq^i=\left[ q^i,G_a\right] \eta _1^a+\left[ q^i,C_a^0\right]
\eta _2^a,
\end{equation}
\begin{equation}
\label{(84)}sp_i=\left[ p_i,G_a\right] \eta _1^a+\left[ p_i,C_a^0\right]
\eta _2^a,
\end{equation}
\begin{equation}
\label{(85)}sz^a=\lambda \eta _2^a,
\end{equation}
\begin{equation}
\label{(86)}s\overline{p}_a=0,
\end{equation}
\begin{equation}
\label{(87)}sv^a=\dot \eta _1^a-C_{bc}^{\quad a}v^b\eta _1^c,
\end{equation}
\begin{equation}
\label{(88)}su^a=\dot \eta _2^a+\left[ C_b^0,f^a\left( C\right) \right] \eta
_2^b-\Delta _{cb}\frac{\partial f^a\left( C\right) }{\partial C_c}\eta _1^b,
\end{equation}
\begin{equation}
\label{(89)}s\eta _1^a=s\eta _2^a=0,
\end{equation}
\begin{equation}
\label{(90)}s\overline{\eta }_1^a=B_1^a,
\end{equation}
\begin{equation}
\label{(91)}s\overline{\eta }_2^a=-B_2^a,
\end{equation}
\begin{equation}
\label{(92)}sB_1^a=sB_2^a=0.
\end{equation}
The path integral correspondent to action (\ref{(82)}) takes the form
\begin{eqnarray}\label{(93)}
 Z_{\Psi ^{\prime }}&=&\int \QTR{cal}{D}q^i\,\QTR{cal}{D}p_i\,\QTR{cal}{D}z^a
\,\QTR{cal}{D}\overline{p}_a\,\QTR{cal}{D}\overline{\eta }_1^a\,\QTR{cal}{D}
\eta _1^a\,\QTR{cal}{D}\overline{\eta }_2^a\,\QTR{cal}{D}\eta _2^a\,
\QTR{cal}{D}v^a\,\QTR{cal}{D}u^a\,\QTR{cal}{D}B_1^a\,\QTR{cal}{D}B_2^a\,
\cdot \nonumber \\
 & &\exp \left( iS_{\Psi ^{\prime }}\right) .
\end{eqnarray} Integrating in (\ref{(93)}) over all the variables excepting
the $q^i$'s and $p_i$'s we derive the following form of the path integral
for the irreducible first-class family
\begin{equation}
\label{(94)}Z_{\Psi ^{\prime }}=\int {\cal D}q^i\,{\cal D}p_i\,\det \left(
\Delta _{cd}\right) \prod\limits_a\delta \left( G_a\right)
\prod\limits_b\delta \left( C_b\right) \exp \left( i\int dt\left( \dot
q^ip_i-\tilde H\right) \right) .
\end{equation}
If we integrate in (\ref{(11)}) over $\mu ^\alpha $'s, it follows
\begin{equation}
\label{(95)}Z_{\Psi ^{\prime }}=Z.
\end{equation}
After performing the above integration, the exponents of the path integrals (%
\ref{(11)}) and (\ref{(94)}) differ through a term which vanishes when $%
C_a=0,$ but this term is not important because of the factors $%
\prod\limits_b\delta \left( C_b\right) $ in the measure from (\ref{(94)}).
It is not hard to see that if we make the transformation (\ref{(5)}) in (\ref
{(94)}) and further integrate over $\left( Q^a,P_a\right) $, we get that $%
Z_{\Psi ^{\prime }}$ will be given by (\ref{indep}). We notify that (\ref
{(94)}) is identical to the path integral derived in \cite{3} in the case of
purely second-class systems. Formula (\ref{(95)}) represents the main result
of this subsection and one of the major results in this paper. It states
that the path integral of the irreducible first-class family is the same
with the one of the original second-class system in our standard gauge (\ref
{(80)}). This is the meaning of applying the BRST quantization to
second-class systems.

\subsection{The quantization in the reducible case}

In this subsection we start with action (\ref{(67)}) and its gauge
invariances (\ref{(68)}-\ref{(77)}). For the sake of generality, we presume
that the first-class family is $k$-th order reducible, with $k\geq 2$. Then,
the ghost spectrum \cite{17} contains the ghosts $\left( \eta _1^a,\eta
_2^a,\eta _3^{a_1},\eta _4^{a_1},\eta _5^{a_1}\right) $, all with the
Grassmann parities and ghost numbers equal to one, as well as the ghosts $%
\eta ^{a_k}$, these ones with the Grassmann parities $k\,\left(
mod\;2\right) $ and ghost numbers $k$. The antifield spectrum contains the
antifields $\left( q_i^{*},p^{*i},z_a^{*},\overline{p}^{*a},y_{a_1}^{*},\pi
^{*a_1},v_a^{*},v_{a_1}^{*},u_a^{*},u_{a_1}^{*}\right) $, all with the
Grassmann parities equal to one and ghost numbers equal to minus one, as
well as the antifields $\eta _{a_{k-1}}^{*}$ with the Grassmann parities $%
k\,\left( mod\;2\right) $ and ghost numbers $\left( -k\right) $. Then, the
minimal solution of the master equation is expressed by%
\begin{eqnarray}\label{(96)}
& &S^{\prime }=S_0^{\prime }\left[ q^i,p_i,z^a,\overline{p}_a,y^{a_1},
\pi _{a_1},v^a,v^{a_1},u^a,u^{a_1}\right] +\int dt\left\{ q_i^{*}\left(
\frac{\partial G_a}{\partial p_i}\eta _1^a+\frac{\partial C_a^0}{\partial p_i}
\eta _2^a-\right. \right. \nonumber \\
& &\left. \frac{\partial Z_{\;a_1}^a}{\partial p_i}\overline{p}_a\eta _4^{a_1}
\right) -p^{*i}\left( \frac{\partial G_a}{\partial q^i}\eta _1^a+\frac{
\partial C_a^0}{\partial q^i}\eta _2^a-\frac{\partial Z_{\;a_1}^a}{\partial
q^i}\overline{p}_a\eta _4^{a_1}\right) +z_a^{*}\left( \lambda \eta _2^a-Z_{
\;a_1}^a\eta _4^{a_1}\right) +\nonumber \\
& &y_{a_1}^{*}\eta _3^{a_1}+v_a^{*}\left( \dot \eta _1^a-C_{bc}^{\quad a}v^b
\eta _1^c\right) +v_{a_1}^{*}\dot \eta _3^{a_1}+u_{a_1}^{*}\left( \dot
\eta _4^{a_1}+\eta _3^{a_1}+\lambda \eta _5^{a_1}\right) +u_a^{*}\left[ \dot
\eta _2^a+\right. \nonumber \\
& &\left[ C_b^0,f^a\left( C\right) \right] \eta _2^b-\Delta _{cb}\frac{
\partial f^a\left( C\right) }{\partial C_c}\eta _1^b-\frac 1\lambda \left( Z_{
\;a_1}^b\left[ C_b^0,f^a\left( C\right) \right] +\left[ Z_{\;a_1}^a,
\tilde H^{*}\right] \right) \eta _4^{a_1}+ \nonumber \\
& &\left. \left. Z_{\;a_1}^a\eta _5^{a_1}\right] +\sum\limits_{j=2}^k
\eta _{a_{j-1}}^{*}Z_{\;a_j}^{a_{j-1}}\eta ^{a_j}+\ldots \right\} ,
\end{eqnarray} where ``$\ldots $'' signify other terms with antighost
numbers greater than one, which are not important due to the concrete form
of the gauge-fixing fermion to be given below. We introduce a non-minimal
sector \cite{17} such that the non-minimal solution of the master equation
to become%
\begin{eqnarray}\label{(97)}
S^{\prime \prime }&=&S^{\prime }+\int dt\left( \overline{\eta }_{1a}^{*}B_1^a+
\overline{\eta }_{2a}^{*}B_2^a+\overline{\eta }_1^{*a_1}B_{1a_1}+\overline{
\eta }_2^{*a_1}B_{2a_1}+\overline{\eta }_3^{*a_1}B_{3a_1}+\right. \nonumber \\
& &\left. \sum\limits_{j=2}^k\overline{\eta }^{*a_j}B_{a_j}\right) .
   \end{eqnarray} The gauge-fixing fermion in the reducible case has the
form
\begin{equation}
\label{(98)}\Psi ^{\prime \prime }=\Psi ^{\prime }+\int dt\left( -\overline{%
\eta }_{1a_1}y^{a_1}-\overline{\eta }_{2a_1}\eta _4^{a_1}-\overline{\eta }%
_{3a_1}u^{a_1}+\sum\limits_{j=2}^k\overline{\eta }_{a_j}\eta ^{a_j}\right) ,
\end{equation}
where $\Psi ^{\prime }$ is given in (\ref{(80)}).The gauge-fixing fermion (%
\ref{(98)}) implements the canonical gauge conditions $C_a=0$, $z^a=0$, $%
y^{a_1}=0$, $u^{a_1}=0$. Eliminating as usually the antifields from (\ref
{(97)}), we infer the following gauge-fixed action%
\begin{eqnarray}\label{(99)}
& &S_{\Psi ^{\prime \prime }}^{\prime \prime }=S_{\Psi ^{\prime }}+\int dt
\left( \overline{\eta }_1^a\Delta _{ab}\frac{\partial f^b\left( \rho \right) }
{\partial \rho _c}\left[ \rho _c,Z_{\;a_1}^d\right] \overline{p}_d\eta _4^{a_1
}-\overline{\eta }_2^a\Delta _{ab}Z_{\;a_1}^b\eta _4^{a_1}-\right. \nonumber \\
& &\overline{\eta }_{1a_1}\eta _3^{a_1}- \frac 1\lambda \overline{\eta }_{3a_1
}\left( \dot \eta _4^{a_1}+\eta _3^{a_1}+\lambda \eta _5^{a_1}\right) -y^{a_1}
B_{1a_1}-\eta _4^{a_1}B_{2a_1}-\nonumber \\
& &\left.\frac 1\lambda u^{a_1}B_{3a_1}+\sum_{j=2}^k\eta ^{a_j}B_{a_j}\right) ,
\end{eqnarray} with $S_{\Psi ^{\prime }}$ as in (\ref{(82)}). Integrating
now in the path integral of the reducible first-class family, $Z_{\Psi
^{\prime \prime }}$ (corresponding to (\ref{(99)})) over all the variables
excepting $q^i$'s and $p_i$'s, we find
\begin{equation}
\label{(100)}Z_{\Psi ^{\prime \prime }}=Z.
\end{equation}
Formula (\ref{(100)}) is the basic result of this subsection and, actually,
of this paper. It expresses the fact that in the reducible case the path
integral of the first-class family is the same with the one of the original
second-class theory. We are able now to explain in what sense the original
second-class system maintains the trace of reducibility of a certain
first-class system. At the classical level, we obtain from (\ref{(67)})
putting all the extravariables equal to zero the Hamiltonian action of the
original system. Thus, at the classical level the second-class system comes
from the reducible first-class family (\ref{(67)}). At the path integral
level, formula (\ref{(100)}) shows that the path integral of the original
system is coming from the BRST quantization of the reducible first-class
family. So, the original system can be regarded at both levels as coming
from the reducible first-class family. This is the meaning of the original
system preserving the relic of the reducibility of the first-class family.

\section{The Lagrangian approach of the first-class family}

In this section we shall derive under some simplifying assumptions the
Lagrangian form of the path integrals deduced in the previous section and
clarify the physical origin of the first-class family in both reducible and
irreducible cases. Related to the physical origin, we shall emphasise the
Wess-Zumino action in these cases. A different way of deriving the
Lagrangian form of the path integral is presented in \cite{22}. There, the
linear part in the Lagrange multipliers associated to the secondary,
tertiary, ... constraints are eliminated through a canonical
transformation.Further, the integration in the path integral over the
momenta and the Lagrange multipliers of the primary constraints leads to a
Lagrangian, while the integration over the remaining multipliers (by
stationary-phase method) gives the Lagrangian measure in the path integral.
In the sequel, we expose an alternative method under special hypotheses.
Again, we shall consider separately the two cases.

\subsection{The Lagrangian approach in the irreducible case}

If in (\ref{(82)}) we make the transformations $f^a\left( \rho \right)
\longrightarrow f^a\left( \rho \right) +\frac 12B_1^a$, which do not affect
its BRST invariances (as $sB_1^a=0$), and integrate in the corresponding
path integral over all the variables excepting $\left( q^i,p_i\right) $, we
find
\begin{eqnarray}\label{(101)}
 Z_{\Psi ^{\prime }}&=&\int \QTR{cal}{D}q^i\,\QTR{cal}{D}p_i\,\prod\limits_e
\delta \left( G_e\right) \det \left( \Delta _{ab}\frac{\partial f^b\left(
C\right) }{\partial C_c}\Delta _{cd}\right) \cdot \nonumber \\
 & &\exp \left( i\int dt\left( \dot q^ip_i-H\right) \right) .
\end{eqnarray} If the $C_a$'s depend only on the $q^i$'s, the integration
over the $p_i$'s in (\ref{(101)}) leads us to the following form of the
Lagrangian path integral
\begin{equation}
\label{(102)}Z_{\Psi ^{\prime }}=\int {\cal D}q^i\,\det \left( \Delta _{ab}
\frac{\partial f^b\left( C\right) }{\partial C_c}\Delta _{cd}\right) \exp
\left( i\int dt\,L_0\left( q^i,\dot q^i\right) \right) ,
\end{equation}
where $L_0\left( q^i,\dot q^i\right) $ is the Lagrangian of the original
second-class system. From (\ref{(102)}) it results that if the original
canonical Hamiltonian is more than quad\-rat\-ic in the functions $C_a$ (see
(\ref{(9)})) in the Lagrangian path integral it will appear the non-trivial
local measure
\begin{equation}
\label{(103)}\mu =\det \left( \Delta _{ab}\frac{\partial f^b\left( C\right)
}{\partial C_c}\Delta _{cd}\right) .
\end{equation}
In the case of $H$ at most quadratic in the $C_a$'s, the measure (\ref{(103)}%
) reduces to $\mu =\det \left( \Delta _{ad}\right) $, so the Lagrangian path
integral takes the simple form
\begin{equation}
\label{(104)}Z_{\Psi ^{\prime }}=\int {\cal D}q^i\,\left( \det C_{\alpha
\beta }\right) ^{1/2}\exp \left( i\int dt\,L_0\left( q^i,\dot q^i\right)
\right) .
\end{equation}
The last formula remains also valid when the $C_a$'s depend on $p_i$'s
because in this case $\left. \frac{\partial ^kH}{\partial C_{a_1}\ldots
\partial C_{a_k}}\right| _{C_a=0}=0$ for any $k>2$.

In the sequel we shall make clear the physical provenance of the irreducible
first-class family. In this end, we consider for simplicity $C_a^0\equiv 0$.
Under this circumstance, action (\ref{(36)}) (which describes the
irreducible first-class family) reduces to
\begin{equation}
\label{(105)}S_0\left[ q^i,p_i,z^a,\overline{p}_a,v^a,u^a\right] =\int
dt\left( \dot q^ip_i+\dot z^a\overline{p}_a-\overline{H}^{*}-v^aG_a-\lambda
u^a\overline{p}_a\right) ,
\end{equation}
where
\begin{equation}
\label{(106)}\overline{H}^{*}=\tilde H-\frac{\lambda ^2}2\Delta ^{ab}
\overline{p}_a\overline{p}_b+\lambda \overline{p}_af^a\left( C\right) .
\end{equation}
Action (\ref{(105)}) takes into account the primary, as well as the
secondary constraints. Passing from this extended action to the total one
\cite{17} (taking $u^a=0$) and making in the resulting action the
transformation (\ref{(5)}), we infer \cite{4}
\begin{eqnarray}\label{(107)}
 & &S_0\left[ Q^a,P_a,z^\Delta ,\overline{p}_\Delta ,z^a,\overline{p}_a,v^a
\right] =\nonumber \\
& &\int dt\left( \dot Q^aP_a+\dot z^\Delta \overline{p}_\Delta +\dot z^a
\overline{p}_a-h^{*}-v^aP_a\right) ,
\end{eqnarray} with
\begin{equation}
\label{(108)}h^{*}=h\left( z^\Delta ,\overline{p}_\Delta \right) -\frac{%
\lambda ^2}2\Delta ^{ab}\overline{p}_a\overline{p}_b+\lambda \overline{p}%
_af^a\left( Q\right) ,
\end{equation}
and $h\left( z^\Delta ,\overline{p}_\Delta \right) =\tilde H\left(
Q^a,z^\Delta ,\overline{p}_\Delta \right) =H\left( 0,z^\Delta ,\overline{p}%
_\Delta \right) $ \cite{4}. Eliminating from (\ref{(107)}) all the momenta
and Lagrangian multipliers on their equations of motion \cite{21}, we get
the Lagrangian action of the irreducible first-class family under the form
\begin{eqnarray}\label{(109)}
& &S_0^L\left[ Q^a,z^\Delta ,z^a\right] =\nonumber \\
& &\int dt\left( l\left( z^\Delta ,\dot z^\Delta \right) -\frac 1{2\lambda ^2}
\Delta _{ab}\left( \lambda f^a\left( Q\right) -\dot z^a\right) \left( \lambda
f^b\left( Q\right) -\dot z^b\right) \right) .
\end{eqnarray} In (\ref{(109)}) $l\left( z^\Delta ,\dot z^\Delta \right) $
is the Lagrangian corresponding to $h\left( z^\Delta ,\overline{p}_\Delta
\right) $. It is clear that for $z^a=0$ action (\ref{(109)}) reduces to the
original Lagrangian action. The gauge invariances of (\ref{(109)}) are as
follows
\begin{equation}
\label{(110)}\delta _\epsilon Q^a=R_{\;b}^a\left( Q\right) \dot \epsilon ^b,
\end{equation}
\begin{equation}
\label{(111)}\delta _\epsilon z^a=\lambda \epsilon ^a,
\end{equation}
\begin{equation}
\label{(112)}\delta _\epsilon z^\Delta =0,
\end{equation}
where $R_{\;b}^a\left( Q\right) $ is the inverse of the matrix $\frac{%
\partial f^a\left( Q\right) }{\partial Q^b}$ (from Theorem 3 it is obvious
that the inverse exists). The gauge transformations (\ref{(110)}-\ref{(112)}%
) result from (\ref{(37)}-\ref{(42)}) via $\delta _\epsilon u^a=0$ \cite{21}.

In order to reveal the origin of the first-class family we consider the
Lagrangian action
\begin{equation}
\label{(113)}\overline{S}_0\left[ z^a\right] =\int dt\left( -\frac
1{2\lambda ^2}\Delta _{ab}\dot z^a\dot z^b\right) .
\end{equation}
This action is invariant under the rigid (Noether) transformations
\begin{equation}
\label{(114)}\delta _\epsilon z^a=\lambda \epsilon ^a,
\end{equation}
with all $\epsilon ^a$ constant. Gauging now symmetries (\ref{(114)}) (i.e. $%
\epsilon ^a$ are arbitrary functions of time), action (\ref{(113)}) is no
more gauge invariant. Thus, it is necessary to introduce in (\ref{(113)})
some additional variables in order to obtain a gauge-invariant action. Under
this observation, it results clearly that action (\ref{(109)}) comes from
the gauging of the rigid symmetries (\ref{(114)}) through the introduction
of the variables $\left( Q^a,z^\Delta \right) $ which transform accordingly (%
\ref{(110)}), (\ref{(112)}). The action of the first-class family contains
some mixing-component terms of the type ``current-current'', with the
``currents''
\begin{equation}
\label{(115)}j_a=\frac 1\lambda \Delta _{ab}\left( \lambda f^b\left(
Q\right) -\dot z^b\right) .
\end{equation}
These ``currents'' are conservative and gauge-invariant and come from the
rigid invariances (\ref{(114)}) of the action (\ref{(109)}) via Noether's
theorem.

The Wess-Zumino action in the irreducible case is defined by%
\begin{eqnarray}\label{(116)}
S_0^{WZ}\left[ Q^a,z^\Delta ,z^a\right] &=&S_0^L\left[ Q^a,z^\Delta ,z^a
\right] -S_0^L\left[ Q^a,z^\Delta ,z^a=0\right] = \nonumber \\
& &\int dt\left( -\frac 1{2\lambda ^2}\Delta _{ab}\dot z^a\dot z^b+\frac 1
\lambda \Delta _{(ab)}f^a\left( Q\right) \dot z^b\right) ,
\end{eqnarray} and obvious vanishes when $z^a=0$. The Wess-Zumino action was
introduced for the first time in the context of anomalous field theories
\cite{18}. For the chiral Schwinger model this action was discovered by
Fadeev and Shatashvili in the framework of the canonical quantization of
this model. In the case of our formalism the Wess-Zumino action is necessary
in order to make gauge-invariant the original second-class system, such that
to apply subsequently the BRST formalism. We remark that a piece in the
Wess-Zumino action is precisely action (\ref{(113)}). From (\ref{(109)}), we
observe that the ``Wess-Zumino variables'' $z^a$ are introduced in order to
compensate in a certain sense the unphysical variables from the original
theory, $Q^a$. Indeed, in (\ref{(109)}) the $z^a$'s are coupled only to the
unphysical variables $Q^a$ through the ``current-current'' terms. The
earlier separation in physical and unphysical variables is a consequence of
the fact that we took $C_a^0\equiv 0$. In the case of field theory it is
exactly the presence of $C_a^0$ non-identically vanishing in $\gamma _a$
which ensures the Lorentz covariance of the Lagrangian action of the
first-class family due to the fact that the function $g$ given in (\ref{(35)}%
) is non-vanishing. The proof of this last conclusion is technically
difficult in general, and this is why we shall exemplify it on the models
exposed in Sec.6.

\subsection{The Lagrangian approach in the reducible case}

The Lagrangian path integral of the reducible first-class family is obtained
analogously with the irreducible case. Making in (\ref{(99)}) the
transformation $f^a\left( \rho \right) \longrightarrow f^a\left( \rho
\right) +\frac 12B_1^a$ and integrating in the correspondent path integral
over all the variables excepting $\left( q^i,p_i\right) $ we get that $%
Z_{\Psi ^{\prime \prime }}$ is also given by formula (\ref{(101)}). The
procedure of passing from (\ref{(101)}) to (\ref{(102)}-\ref{(104)}) is
identical with the one from the irreducible case, finally deriving the same
results.

Next, we shall analyze the origin of the reducible first-class family. We
shall consider the case $C_a^0=0$, too. Now, action (\ref{(67)}) takes the
form%
\begin{eqnarray}\label{(117)}
& &S_0^{\prime }\left[ q^i,p_i,z^a,\overline{p}_a,y^{a_1},\pi _{a_1},v^a,
v^{a_1},u^a,u^{a_1}\right] =\int dt\left( \dot q^ip_i+\right. \nonumber \\
& &\left. \dot z^a\overline{p}_a+\dot y^{a_1}\pi _{a_1}-\widehat{H^{*}}-
v^aG_a-v^{a_1}\gamma _{a_1}-\lambda u^a\overline{p}_a-u^{a_1}\overline{\gamma
}_{a_1}\right) ,
\end{eqnarray} where $\widehat{H^{*}}=\overline{H}^{*}-y^{a_1}\overline{%
\gamma }_{a_1}$. The passing from the extended action (\ref{(117)}) to its
correspondent total action is gained putting $u^a=u^{a_1}=0$. Making in this
total action the transformation (\ref{(5)}), we get%
\begin{eqnarray}\label{(118)}
& &S_0^{\prime }\left[ Q^a,P_a,z^\Delta ,\overline{p}_\Delta ,z^a,
\overline{p}_a,y^{a_1},\pi _{a_1},v^a,v^{a_1}\right] =\int dt\left( \dot
Q^aP_a+\right. \nonumber \\
& &\left. \dot z^\Delta \overline{p}_\Delta +\dot z^a\overline{p}_a+
\dot y^{a_1}\pi _{a_1}-h^{*}+y^{a_1}\overline{\gamma }_{a_1}-v^aP_a-v^{a_1}
\gamma _{a_1}\right) .
\end{eqnarray} Eliminating from action (\ref{(118)}) the momenta and the
Lagrangian multipliers on their equations of motion, we deduce the
Lagrangian action of the reducible first-class family as%
\begin{eqnarray}\label{(119)}
& &S_0^{\prime L}\left[ Q^a,z^\Delta ,z^a,y^{a_1}\right] =\int dt\left( l
\left( z^\Delta ,\dot z^\Delta \right) - \right. \nonumber \\
& &\left. \frac 1{2\lambda ^2}\Delta _{ab}\left( \lambda f^a\left( Q\right) +
Z_{\;a_1}^ay^{a_1}-\dot z^a\right) \left( \lambda f^b\left( Q\right) +
Z_{\;b_1}^by^{b_1}-\dot z^b\right) \right) .
\end{eqnarray} In order to obtain action (\ref{(119)}) we presumed that $%
Z_{\;a_1}^a$'s do not depend on the momenta. Action (\ref{(119)}) is
invariant under the gauge transformations
\begin{equation}
\label{(120)}\delta _\epsilon Q^a=R_{\;b}^a\left( Q\right) \left( \dot
\epsilon ^b+Z_{\;b_1}^b\overline{\epsilon }^{b_1}-\dot Z_{\;b_1}^b\epsilon
^{b_1}\right) ,
\end{equation}
\begin{equation}
\label{(121)}\delta _\epsilon z^a=\lambda \epsilon ^a-Z_{\;a_1}^a\epsilon
^{a_1},
\end{equation}
\begin{equation}
\label{(122)}\delta _\epsilon y^{a_1}=-\dot \epsilon ^{a_1}-\lambda
\overline{\epsilon }^{a_1},
\end{equation}
\begin{equation}
\label{(123)}\delta _\epsilon z^\Delta =0.
\end{equation}
The gauge transformations (\ref{(120)}-\ref{(123)}) result from (\ref{(68)}-%
\ref{(77)}) via $\delta _\epsilon u^a=\delta _\epsilon u^{a_1}=0$.Due to (%
\ref{(44)}) it follows that $Z_{\;a_1}^a$ do not depend on the $Q^a$'s, such
that, from (\ref{(123)}) we have $\delta _\epsilon Z_{\;a_1}^a=0$. As
expected, (\ref{(120)}-\ref{(123)}) represent a set of reducible gauge
transformations. If the functions $C_a^0$ are $k$-th order reducible, then (%
\ref{(120)}-\ref{(123)}) possess the same reducibility order. Indeed,
denoting $X^\alpha =\left( Q^a,z^\Delta ,z^a,y^{a_1}\right) $ we have the
reducibility relations
\begin{equation}
\label{(124)}Z_{\;\alpha }^a\delta _\epsilon X^\alpha =0,
\end{equation}
with
\begin{equation}
\label{(125)}Z_{\;\alpha }^a=\left( \lambda \delta _{\;b}^a,0,\dot
R_{\;b}^a,R_{\;b}^aZ_{\;a_1}^b\right) .
\end{equation}
Obviously, the reducibility relations (\ref{(124)}) are written in De Witt
notations. Because of (\ref{(50)}-\ref{(51)}) we further find the
reducibility relations
\begin{equation}
\label{(126)}Z_{\;a}^{A_1}Z_{\;\alpha }^a=0,
\end{equation}
\begin{equation}
\label{(127)}Z_{\;A_1}^{A_2}Z_{\;a}^{A_1}=0,
\end{equation}
$$
\vdots
$$
\begin{equation}
\label{(128)}Z_{\;A_{k-1}}^{A_{k-2}}Z_{\;A_k}^{A_{k-1}}=0,
\end{equation}
where $Z_{\;a}^{A_1}=\left( 0,0,0,Z_{\;a_2}^{a_1}\right) $, and $%
Z_{\;A_{k-j+1}}^{A_{k-j}}=\left( 0,0,0,Z_{\;a_{k-j+1}}^{a_{k-j}}\right) $.

In the reducible case, too, action (\ref{(119)}) results from the gauging of
some rigid symmetries. In this case, there appear two important cases.

i) Firstly, we consider
\begin{equation}
\label{(129)}Z_{\;a_1}^a\Theta ^{a_1}\neq 0,
\end{equation}
for\ all\ $\Theta ^{a_1}$'s\ constant, (i.e. the $Z_{\;a_1}^a$'s do not
contain derivative terms to act upon the $\Theta ^{a_1}$'s). Next, we shall
show that action (\ref{(119)}) also comes from the gauging of some rigid
symmetries of action (\ref{(113)}). Action (\ref{(113)}) is invariant under
the rigid transformations (with two sets of constant parameters)
\begin{equation}
\label{(130)}\delta _\epsilon z^a=\lambda \epsilon ^a-Z_{\;a_1}^a\epsilon
^{a_1}.
\end{equation}
The gauging of the last symmetries implies the necessity of introducing some
new variables in order to obtain from (\ref{(113)}) a gauge-invariant
action. As $\delta _\epsilon \overline{S}_0=\int dt\left( -\frac 1{\lambda
^2}\Delta _{(ab)}\dot z^b\left( \lambda \dot \epsilon ^a-Z_{\;a_1}^a\dot
\epsilon ^{a_1}-\dot Z_{\;a_1}^a\epsilon ^{a_1}\right) \right) $ it follows
that it is necessary to introduce the variables $\left( Q^a,z^\Delta
,y^{a_1}\right) $ having the gauge transformations
\begin{equation}
\label{(131)}\delta _\epsilon Q^a=R_{\;b}^a\left( Q\right) \left( \dot
\epsilon ^b-\dot Z_{\;b_1}^b\epsilon ^{b_1}\right) ,
\end{equation}
\begin{equation}
\label{(132)}\delta _\epsilon y^{a_1}=-\dot \epsilon ^{a_1},
\end{equation}
\begin{equation}
\label{(133)}\delta _\epsilon z^\Delta =0,
\end{equation}
such that the gauge-invariant action deriving from $\overline{S}_0$ to have
precisely the form (\ref{(119)}). We notice that the introduction of the
terms $\lambda f^a\left( Q\right) +Z_{\;a_1}^ay^{a_1}$ in $\overline{S}_0$
(in order to get (\ref{(119)})) allows the additional gauge invariances of
this term of the form
\begin{equation}
\label{(134)}\delta _\epsilon Q^a=R_{\;b}^a\left( Q\right) Z_{\;b_1}^b
\overline{\epsilon }^{b_1},
\end{equation}
\begin{equation}
\label{(135)}\delta _\epsilon y^{a_1}=-\lambda \overline{\epsilon }^{a_1},
\end{equation}
which are due to the manifest reducibility of the first-class family. The
gauge invariances (\ref{(131)}-\ref{(135)}) are nothing but (\ref{(120)}-\ref
{(123)}). In this way we evidenced that action (\ref{(119)}) comes from the
gauging of the rigid symmetries (\ref{(130)}) of action (\ref{(113)}). At
the same time, action (\ref{(119)}) is invariant under the rigid
transformations (\ref{(130)}). Then, there result from Noether's theorem the
conserved ``currents''
\begin{equation}
\label{(136)}j_a=\frac 1\lambda \Delta _{ab}\left( \lambda f^b\left(
Q\right) +Z_{\;b_1}^by^{b_1}-\dot z^b\right) ,
\end{equation}
corresponding to the rigid parameters $\epsilon ^a$, and
\begin{equation}
\label{(137)}j_{a_1}=-\frac 1{\lambda ^2}Z_{\;a_1}^a\Delta _{ab}\left(
\lambda f^b\left( Q\right) +Z_{\;b_1}^by^{b_1}-\dot z^b\right) \equiv
-\lambda Z_{\;a_1}^aj_a,
\end{equation}
associated to the rigid parameters $\epsilon ^{a_1}$. These ``currents''
present an interesting feature, namely they are $k$-th order reducible.
Using (\ref{(136)}-\ref{(137)}) we have the reducibility relations
\begin{equation}
\label{(138)}Z_{\;a_1}^aj_a+\lambda j_{a_1}\equiv Z_{\;a_1}^\Lambda
j_\Lambda =0,
\end{equation}
where $Z_{\;a_1}^\Lambda =\left( Z_{\;a_1}^a,\lambda \delta
_{\;a_1}^{b_1}\right) $ and $j_\Lambda =\left( j_a,j_{b_1}\right) $. From (%
\ref{(50)}-\ref{(51)}) we further find the reducibility relations
\begin{equation}
\label{(139)}Z_{\;\Lambda _1}^{a_1}Z_{\;a_1}^\Lambda =0,
\end{equation}
\begin{equation}
\label{(140)}Z_{\;\Lambda _2}^{\Lambda _1}Z_{\;\Lambda _1}^{a_1}=0,
\end{equation}
$$
\vdots
$$
\begin{equation}
\label{(141)}Z_{\;\Lambda _k}^{\Lambda _{k-1}}Z_{\;\Lambda _{k-1}}^{\Lambda
_{k-2}}=0,
\end{equation}
with $Z_{\;\Lambda _1}^{a_1}=\left( Z_{\;a_2}^{a_1},0\right) $ and $%
Z_{\;\Lambda _{k-j}}^{\Lambda _{k-j-1}}=\left(
Z_{\;a_{k-j}}^{a_{k-j-1}},0\right) $. The reducible ``currents'' (\ref{(136)}%
-\ref{(137)}) are gauge-invariant under the gauge transformations (\ref
{(120)}-\ref{(123)}). Thus, the action of the reducible first-class family
contains some mixing-component terms of the type ``current-current'' $-\frac
12\Delta ^{ab}j_aj_b$, with $j_a$ given by (\ref{(136)}).

ii) Secondly, we consider
\begin{equation}
\label{(142)}Z_{\;a_1}^a\Theta ^{a_1}=0,
\end{equation}
only for\ all\ $\Theta ^{a_1}$'s\ constant, (i.e. the $Z_{\;a_1}^a$'s
contain derivative terms to act upon the $\Theta ^{a_1}$'s). Now we prove
that action (\ref{(119)}) also results from the gauging of some rigid
symmetries, but not for action (\ref{(113)}). We start with the action
\begin{equation}
\label{(143)}\tilde S_0\left[ z^a,y^{a_1}\right] =-\int dt\left( \frac
1{2\lambda ^2}\Delta _{ab}\left( Z_{\;a_1}^ay^{a_1}-\dot z^a\right) \left(
Z_{\;b_1}^by^{b_1}-\dot z^b\right) \right) ,
\end{equation}
which is invariant under the rigid transformations
\begin{equation}
\label{(144)}\delta _\epsilon z^a=\lambda \epsilon ^a,
\end{equation}
\begin{equation}
\label{(145)}\delta _\epsilon y^{a_1}=-\lambda \overline{\epsilon }^{a_1}.
\end{equation}
Gauging these symmetries, we infer
$$
\delta _\epsilon \tilde S_0=\int dt\frac 1\lambda \Delta _{(ab)}\left(
Z_{\;a_1}^a\overline{\epsilon }^{a_1}+\dot \epsilon ^a\right) \left(
Z_{\;b_1}^by^{b_1}-\dot z^b\right) ,
$$
so that it is necessary to introduce the new variables $\left( Q^a,z^\Delta
\right) $ with the gauge transformations
\begin{equation}
\label{new}\delta _\epsilon Q^a=R_{\;b}^a\left( Q\right) \left( \dot
\epsilon ^b+Z_{\;b_1}^b\overline{\epsilon }^{b_1}\right) ,
\end{equation}
and (\ref{(123)}) for the $z^\Delta $'s, further resulting the
gauge-invariant action derived from $\tilde S_0$ precisely of the form (\ref
{(119)}). We observe that action (\ref{(143)}) possesses the additional
gauge invariances
\begin{equation}
\label{invg}\delta _\epsilon Q^a=-\frac 1\lambda R_{\;b}^a\left( Q\right)
\dot Z_{\;b_1}^b\epsilon ^{b_1},
\end{equation}
\begin{equation}
\label{(146)}\delta _\epsilon z^a=-Z_{\;a_1}^a\epsilon ^{a_1},
\end{equation}
\begin{equation}
\label{(147)}\delta _\epsilon y^{a_1}=-\dot \epsilon ^{a_1},
\end{equation}
such that $\delta _\epsilon \left( \lambda f^a\left( Q\right)
+Z_{\;a_1}^ay^{a_1}-\dot z^a\right) =0$ under the prior transformations.
Thus, the last transformations are independent of the non-invariant form of (%
\ref{(143)}) under the gauge version of (\ref{(144)}-\ref{(145)}). In the
case of $\dot Z_{\;b_1}^b=0$, the invariances (\ref{invg}-\ref{(147)})
reduce to (\ref{(146)}-\ref{(147)}), the last ones representing some gauge
symmetries characteristic to the terms $\left( Z_{\;a_1}^ay^{a_1}-\dot
z^a\right) $ containing only extravariables. To conclude with, action (\ref
{(119)}) comes from the gauging of the rigid symmetries (\ref{(144)}-\ref
{(145)}) of action (\ref{(143)}) and, in the same time, possesses some
supplementary gauge invariances because (\ref{(142)}) do not hold if $\Theta
^{a_1}$'s are functions of time. Obviously, action (\ref{(119)}) is also
invariant under (\ref{(144)}-\ref{(145)}) so that we obtain via Noether's
theorem the gauge-invariant ``currents'' (\ref{(136)}) (the rigid symmetries
(\ref{(145)}) lead to some trivial ``currents'') which are no longer
reducible.

In the end of this subsection we emphasise the Wess-Zumino action
corresponding to the reducible case%
\begin{eqnarray}\label{(148)}
S_0^{\prime WZ}\left[ Q^a,z^\Delta ,z^a,y^{a_1}\right] &=&S_0^{\prime }\left[
Q^a,z^\Delta ,z^a,y^{a_1}\right] -S_0^{\prime }\left[ Q^a,z^\Delta ,z^a=0,
y^{a_1}=0\right] \equiv \nonumber \\
& &-\int dt\left( \frac 1{2\lambda ^2}\Delta _{ab}\left( Z_{\;a_1}^ay^{a_1}-
\dot z^a\right) \left( Z_{\;b_1}^by^{b_1}-\dot z^b\right) +\right. \nonumber \\
& &\left. \frac 1\lambda \Delta _{(ab)}f^a\left( Q\right) \left( Z_{\;b_1}^by^
{b_1}-\dot z^b\right) \right) .
\end{eqnarray}
We remark that in the reducible case the Wess-Zumino action contains action (%
\ref{(143)}).

\section{Examples}

In this section we illustrate the general theory presented in this paper on
two representative models.

\subsection{The Massive Yang-Mills theory}

The Lagrangian action describing the massive Yang-Mills theory reads
\begin{equation}
\label{(149)}S_0^L\left[ A\right] =\int d^4x\left( -\frac 14F_{\mu \nu
}^aF_a^{\mu \nu }-\frac 12M^2A_\mu ^aA_a^\mu \right) ,
\end{equation}
where $F_{\mu \nu }^a=\partial _\mu A_\nu ^a-\partial _\nu A_\mu
^a-f_{\;bc}^aA_\mu ^bA_\nu ^c$. The canonical analysis of this model
furnishes the canonical Hamiltonian
\begin{equation}
\label{(150)}H=\int d^3x\left( \frac 12\pi _{ia}\pi _i^a+\frac
14F_{ij}^aF_a^{ij}-A_0^aD_i\pi _a^i+\frac 12M^2A_\mu ^aA_a^\mu \right) ,
\end{equation}
together with the primary, respectively secondary constraints
\begin{equation}
\label{(151)}G_a\equiv \pi _a^0=0,
\end{equation}
\begin{equation}
\label{(152)}C_a\equiv -D_i\pi _a^i+M^2A_a^0=0,
\end{equation}
where $D_i\pi _a^i=\partial _i\pi _a^i-f_{\;ab}^c\pi _c^iA_i^b$ and $\pi
_a^\mu $ denote the canonical momenta of $A_\mu ^a$. It is easy to see that
the above constraints are second-class as the matrix $\left[ C_a,G_b\right]
=M^2\delta _{ab}\equiv \Delta _{ab}$ has a non-vanishing determinant. We
observe that $\left[ G_a,G_b\right] =0$ strongly. We choose
\begin{equation}
\label{(153)}C_a^0=-\partial _i\pi _a^i,
\end{equation}
and
\begin{equation}
\label{(154)}C_a^1=f_{\;ab}^c\pi _c^iA_i^b+M^2A_a^0,
\end{equation}
such that $\left[ C_a^0,C_b^0\right] =\left[ C_a^0,G_b\right] =0$, and $%
\left[ C_a^1,G_b\right] =\Delta _{ab}$. It follows that the massive
Yang-Mills theory verifies the hypotheses of our methods. The Hamiltonian $%
H^{\prime }$ for our model is given by
\begin{eqnarray}\label{(155)}
& &H^{\prime }=\int d^3x\left( \frac 12\pi _{ia}\pi _i^a+\frac 14F_{ij}^aF_a^{
ij}+\frac 12M^2A_i^aA_a^i-\right. \nonumber \\
& &\left. \frac 1{2M^2}f_{\;ba}^cf_{\;de}^a\pi _c^iA_i^b\pi _j^dA^{je}\right) ,
\end{eqnarray} such that $\left[ H^{\prime },G_a\right] =0$. The functions $%
C_a^0$ being irreducible, we introduce the additional canonical pairs $%
\left( \varphi ^a,\Pi _a\right) $ in number equal to the number of pairs $%
\left( G_a,C_a\right) $. The above canonical pairs play the role of the
pairs $\left( z^a,\overline{p}_a\right) $ from the general theory. The
functions $f^a\left( C\right) $ for our model have the form
\begin{equation}
\label{(156)}f^a\left( C\right) =\Delta ^{ab}C_b.
\end{equation}
The Hamiltonian of the first class family is
\begin{eqnarray}\label{(157)}
 & &H^{*}=H^{\prime }+\int d^3x\left( -\frac{\lambda ^2}{2M^2}\Pi _a\Pi ^a-
\right. \nonumber \\
 & &\left.\frac 1{M^2}\left( \lambda \Pi _a-\partial _i\pi _a^i\right) \left(
f_{\;bc}^a\pi _j^bA^{jc}-M^2A_0^a\right) +g\right) ,
\end{eqnarray} where
\begin{eqnarray}\label{(158)}
& &g\left( A,\pi ,\varphi ,\Pi \right) = \nonumber \\
& &\int d^3x\left( \frac 1{M^2}f_{\,\,bc}^a\Pi _a\pi _i^b\partial ^i\varphi ^c
+\frac 1\lambda \left( f_{abc}\partial _iA_j^a\left( A^{ib}\partial ^j
\varphi ^c-A^{jb}\partial ^i\varphi ^c\right) -\right. \right. \nonumber \\
& &\left. f_{\,\,mn}^af_{abc}A_i^mA_j^nA^{ib}\partial ^j\varphi ^c+\frac 1{M^2
}f_{\,\,\,bc}^af_{amn\text{\thinspace }}\pi ^{ic}\pi _j^mA_i^b\partial ^j
\varphi ^n-M^2A_i^a\partial ^i\varphi _a\right) +\nonumber \\
& &\,\frac 1{\lambda ^2}\left( \frac 12M^2\partial ^i\varphi ^a\partial _i
\varphi _a-\frac 1{2M^2}f_{abc}f_{\,\,mn}^c\pi _i^a\pi _j^m\partial ^i\varphi
^b\partial ^j\varphi ^n-f_{abc}\partial _iA_j^a\partial ^i\varphi ^b\partial
^j\varphi ^c+\right. \nonumber \\
& &\left. \,\frac 12f_{\,\,bc}^af_{amn}A_i^b\partial _j\varphi ^c\left( A^{im}
\partial ^j\varphi ^n-A^{jm}\partial ^i\varphi ^n\right) +\frac 12f_{\,\,mn
}^af_{abc}A_i^mA_j^n\partial ^i\varphi ^b\partial ^j\varphi ^c\right) -
\nonumber \\
& &\left. \frac 1{\lambda ^3}f_{\,\,bc}^af_{amn}A_i^m\partial ^i\varphi ^b
\partial _j\varphi ^c\partial ^j\varphi ^n+\frac 1{4\lambda ^4}f_{\,\,bc}^af_
{amn}\partial ^i\varphi ^b\partial ^j\varphi ^c\partial _i\varphi ^m\partial
_j\varphi ^n\right) .
\end{eqnarray} The first-class constraints of the first-class family are $%
G_a\equiv \pi _a^0=0$ and $\gamma _a\equiv \lambda \Pi _a-\partial _i\pi
_a^i=0$, such that the gauge algebra of the first-class family reads
\begin{equation}
\label{(159)}\left[ G_a,G_b\right] =\left[ G_a,\gamma _b\right] =\left[
\gamma _a,\gamma _b\right] =0,
\end{equation}
\begin{equation}
\label{(160)}\left[ G_a,H^{*}\right] =-\gamma _a,
\end{equation}
\begin{equation}
\label{(161)}\left[ \gamma _a,H^{*}\right] =0.
\end{equation}
The gauge invariances of the extended action are in this case: $\delta
_\epsilon A_0^a=\epsilon _1^a$, $\delta _\epsilon A_i^a=\partial _i\epsilon
_2^a$, $\delta _\epsilon \pi _a^\mu =0$, $\delta _\epsilon \varphi
^a=\lambda \epsilon _2^a$, $\delta _\epsilon \Pi _a=0$, $\delta _\epsilon
v^a=\dot \epsilon _1^a$, $\delta _\epsilon u^a=\dot \epsilon _2^a-\epsilon
_1^a$. The gauge-fixing fermion (\ref{(80)}) for our model reads
\begin{equation}
\label{(162)}\Psi ^{\prime }=-\int d^4x\left( \overline{\eta }_1^a\left(
M^2A_a^0-\frac 1\lambda f_{\;ac}^b\pi _b^i\tilde A_i^c\right) +\lambda M^2
\overline{\eta }_2^a\varphi ^a\right) ,
\end{equation}
with $\tilde A_\mu ^a=A_\mu ^a-\frac 1\lambda \partial _\mu \varphi ^a$. The
path integral (\ref{(94)}) for the massive Yang-Mills theory in the gauge (%
\ref{(162)}) after integration over $\left( A_0^a,\pi _a^0\right) $ is given
by
\begin{equation}
\label{(163)}Z_{\Psi ^{\prime }}=\int {\cal D}A_i^a\,{\cal D}\pi _a^i\,\exp
\left( i\overline{S}\right) ,
\end{equation}
where
\begin{eqnarray}\label{(164)}
& &\overline{S}=\int d^4x\left( \dot A_i^a\pi _a^i-\frac 12\pi _{ia}\pi _i^a-
\frac 14F_{ij}^aF_a^{ij}-\right. \nonumber \\
& &\left. \frac 12M^2A_i^aA_a^i+\frac 1{2M^2}\left( D_i\pi _a^i\right) ^2
\right) .
\end{eqnarray} The results obtained in (\ref{(163)}-\ref{(164)}) are
identical to the ones derived in \cite{2}, \cite{4} through other methods.

The gauge invariances of the total action for our model are inferred from
the ones of the extended action taking $\delta _\epsilon u^a=0$, which
further implies $\dot \epsilon _2^a=\epsilon _1^a$. Then, the gauge
invariances of the total action are given by: $\delta _\epsilon A_\mu
^a=\partial _\mu \epsilon _2^a$, $\delta _\epsilon \pi _a^\mu =0$, $\delta
_\epsilon \varphi ^a=\lambda \epsilon _2^a$, $\delta _\epsilon \Pi _a=0$, $%
\delta _\epsilon v^a=\ddot \epsilon _2^a$. These gauge transformations are
written now under a manifestly covariant form. This is because of the
non-trivial term $-\partial _i\pi _a^i\equiv C_a^0$ in the secondary
constraints $\gamma _a$. It is precisely this term which induces $\delta
_\epsilon A_i^a=\partial _i\epsilon _2^a\neq 0$ and so further implies the
above covariance of the gauge transformations and also $g\left( A,\pi
,\varphi ,\Pi \right) \neq 0$. We shall see below that $g\left( A,\pi
,\varphi ,\Pi \right) $ of the form (\ref{(158)}) ensures the manifestly
covariance of the Lagrangian action for the first-class family. We can reach
this action eliminating from the total action the momenta and Lagrange
multipliers $v^a$ on their equations of motion, namely
\begin{equation}
\label{(165)}\pi _a^i=-\tilde F_a^{0i},
\end{equation}
\begin{equation}
\label{(166)}\Pi _a=\frac 1\lambda \left( M^2\tilde A_a^0+f_{abc}\tilde
F_{0i}^b\tilde A^{ic}\right) ,
\end{equation}
\begin{equation}
\label{(167)}\pi _a^0=0,
\end{equation}
with $\tilde F_{\mu \nu }^a=\partial _\mu \tilde A_\nu ^a-\partial _\nu
\tilde A_\mu ^a-f_{\;bc}^a\tilde A_\mu ^b\tilde A_\nu ^c$. Then, the
Lagrangian action of the first-class family reads
\begin{equation}
\label{(168)}S_0^L\left[ A,\varphi \right] =\int d^4x\left( -\frac 14\tilde
F_{\mu \nu }^a\tilde F_a^{\mu \nu }-\frac 12M^2\tilde A_\mu ^a\tilde A_a^\mu
\right) .
\end{equation}
This action has the gauge invariances $\delta _\epsilon A_\mu ^a=\partial
_\mu \epsilon _2^a$, $\delta _\epsilon \varphi ^a=\lambda \epsilon _2^a$ and
comes from the gauging of the action
\begin{equation}
\label{(169)}S_0\left[ \varphi \right] =-\int d^4x\,\frac{M^2}{2\lambda ^2}%
\partial _\mu \varphi ^a\partial ^\mu \varphi _a,
\end{equation}
which allows the rigid symmetries $\delta _\epsilon \varphi ^a=\lambda
\epsilon _2^a$, with $\epsilon _2^a$ all constant. The conserved
gauge-invariant currents corresponding to the last rigid invariance, but for
action (\ref{(168)}) are
\begin{equation}
\label{(170)}j_a^\mu =\frac 1\lambda \left( f_{\;ac}^b\tilde A_\nu ^c\tilde
F_b^{\mu \nu }-M^2\tilde A_a^\mu \right) .
\end{equation}
Action (\ref{(168)}) together with the currents (\ref{(170)}) coincide in
the abelian limit ($f_{\;bc}^a=0$) with our results derived in \cite{24}and
also with the one resulting from Stueckelberg's formalism \cite{25}. The
Wess-Zumino action in this case takes the form
\begin{eqnarray}\label{(171)}
 & &S_0^{WZ}\left[ A,\varphi \right] =-\frac 1{2\lambda }\int d^4x\left( f_{\;
bc}^a\left( A_\mu ^b\partial _\nu \varphi ^c+A_\nu ^c\partial _\mu \varphi ^b-
\frac 1\lambda \partial _\mu \varphi ^b\partial _\nu \varphi ^c\right) \cdot
\right. \nonumber \\
 & &\left( F_a^{\mu \nu }+\frac 1{2\lambda }g^{\mu \alpha }g^{\nu \beta }f_
{ade}\left( A_\alpha ^d\partial _\beta \varphi ^e+A_\beta ^e\partial _\alpha
\varphi ^d-\frac 1\lambda \partial _\alpha \varphi ^d\partial _\beta
\varphi ^e\right) \right) +\nonumber \\
 & &\left. M^2\left( \frac 1\lambda \partial _\mu \varphi ^a\partial ^\mu
\varphi _a-2A_\mu ^a\partial ^\mu \varphi _a\right) \right) .
\end{eqnarray} This ends the analysis of the model under consideration.

\subsection{Massive abelian three-form gauge fields}

This model is an example of reducible theory. We are starting with the
Lagrangian action \cite{26}
\begin{equation}
\label{(172)}S_0^L\left[ A\right] =\int d^4x\left( -\,\frac 1{2\cdot
4!}F_{\alpha \beta \gamma \rho }F^{\alpha \beta \gamma \rho }-\frac{M^2}{%
2\cdot 3!}A_{\alpha \beta \gamma }A^{\alpha \beta \gamma }\right) ,
\end{equation}
where $F^{\alpha \beta \gamma \rho }=\partial ^{[\alpha }A^{\beta \gamma
\rho ]}\equiv \partial ^\alpha A^{\beta \gamma \rho }-\partial ^\beta
A^{\alpha \gamma \rho }+\partial ^\gamma A^{\rho \alpha \beta }-\partial
^\rho A^{\gamma \alpha \beta }$, with $A^{\alpha \beta \gamma }$'s
antisymmetric in all indices. The canonical analysis of this model provides
the canonical Hamiltonian
\begin{equation}
\label{(173)}H=\int d^3x\left( 3\pi _{ijk}\pi _{ijk}-3A_{0jk}\partial _i\pi
^{ijk}+\frac{M^2}{2\cdot 3!}A_{\alpha \beta \gamma }A^{\alpha \beta \gamma
}\right) ,
\end{equation}
as well as the primary, respectively secondary constraints
\begin{equation}
\label{(174)}G^{ij}\equiv \pi ^{0ij}=0,
\end{equation}
\begin{equation}
\label{(175)}C^{ij}\equiv -3\partial _k\pi ^{kij}+\frac{M^2}2A^{0ij}=0,
\end{equation}
where $\pi ^{kij}$'s are the canonical momenta of the $A_{kij}$'s. It is
simple to check that the above constraints are second-class, the matrix $%
\left[ G_{ij},C_{kl}\right] \equiv \Delta _{ij;kl}=-\frac{M^2}4\left(
g_{ik}g_{jl}-g_{il}g_{jk}\right) $ having a non-vanishing determinant. In
the last relation $g_{ij}$'s denote the spatial part of the Minkowskian
metric. It is clear that $\left[ G_{ij},G_{kl}\right] =0$ strongly. We take
the analogous of $C_a^0$'s, respectively $C_a^1$'s of the form
\begin{equation}
\label{(176)}C_{ij}^0\equiv -3\partial ^k\pi _{kij},
\end{equation}
\begin{equation}
\label{(177)}C_{ij}^1=\frac{M^2}2A_{0ij},
\end{equation}
so the model satisfies the conditions from our methods. In this case, the
functions $C_{ij}^0$ are second-order reducible, the reducibility relations
being $\partial ^iC_{ij}^0=0$ and $\partial ^j\partial ^iC_{ij}^0=0$. The
Hamiltonian $H^{\prime }$ is in the present case
\begin{equation}
\label{(178)}H^{\prime }=\int d^3x\left( 3\pi _{ijk}\pi _{ijk}+\frac{M^2}{%
2\cdot 3!}A_{ijk}A^{ijk}\right) .
\end{equation}
Because we are in the reducible case, we extend the original phase-space as
follows. For every pair $\left( G_{ij},C_{ij}\right) $ we introduce a
bosonic canonical pair $\left( A^{ij},\Pi _{ij}\right) $, with the new
fields antisymmetric in their indices, such that the new secondary
constraints to be
\begin{equation}
\label{(179)}\gamma _{ij}\equiv \lambda \Pi _{ij}-3\partial ^k\pi _{kij}=0.
\end{equation}
It is simply to check that
\begin{equation}
\label{(180)}\partial ^i\gamma _{ij}=\lambda \partial ^i\Pi _{ij}=0.
\end{equation}
For every relation (\ref{(180)}) we introduce the new canonical pair $\left(
A^{0i},\Pi _{0i}\right) $ and the constraint
\begin{equation}
\label{(181)}\gamma _i\equiv \Pi _{0i}=0,
\end{equation}
such that its consistency to imply the constraint
\begin{equation}
\label{(182)}\overline{\gamma }_i=-\partial ^j\Pi _{ji}=0.
\end{equation}
In this way we associated to the original system a one-parameter family of
first-class systems with the first class constraints (\ref{(174)}), (\ref
{(179)}), (\ref{(181)}-\ref{(182)}). Now, the first-class constraints become
reducible
\begin{equation}
\label{(183)}\lambda \overline{\gamma }_j+\partial ^i\gamma _{ij}=0.
\end{equation}
The first-class Hamiltonian of the reducible first-class family reads%
$$
H^{*}=H^{\prime }+\int d^3x\left( -\lambda ^2\Pi _{ij}\Pi
^{ij}+A^{0ij}\gamma _{ij}-2A^{0j}\partial ^i\Pi _{ij}+\right.
$$
\begin{equation}
\label{(184)}\left. +\frac{M^2}{3!\lambda ^2}\left( \frac 12\partial
^{[i}A^{jk]}\partial _{[i}A_{jk]}-A^{ijk}\partial _{[i}A_{jk]}\right)
\right) ,
\end{equation}
where
\begin{equation}
\label{(185)}g=\int d^3x+\frac{M^2}{3!\lambda ^2}\left( \frac 12\partial
^{[i}A^{jk]}\partial _{[i}A_{jk]}-A^{ijk}\partial _{[i}A_{jk]}\right) .
\end{equation}
The gauge invariances of the extended action for this model are $\delta
_\epsilon A^{0ij}=\epsilon _1^{ij}$, $\delta _\epsilon A^{ijk}=\partial
_{}^{[i}\epsilon _2^{jk]}$, $\delta _\epsilon A^{0i}=\epsilon _1^{0i}$, $%
\delta _\epsilon A^{ij}=\lambda \epsilon _2^{ij}+\frac 12\partial
_{}^{[i}\epsilon _2^{0j]}$, $\delta _\epsilon \pi _{0ij}=\delta _\epsilon
\pi _{ijk}=\delta _\epsilon \Pi _{0i}=\delta _\epsilon \Pi _{ij}=0$, $\delta
_\epsilon v^{ij}=\dot \epsilon _1^{ij}$, $\delta _\epsilon v^i=\dot \epsilon
_1^{0i}$, $\delta _\epsilon u^{ij}=\dot \epsilon _2^{ij}-\epsilon
_1^{ij}-\frac 12\partial ^{[i}\epsilon ^{j]}$, $\delta _\epsilon u^j=\dot
\epsilon _2^{0j}-2\epsilon _1^{0j}+\lambda \epsilon ^j$, with $\epsilon ^j$
due to the reducibility relations (\ref{(183)}) (they play the role of $%
\epsilon _5^{a_1}$ in the general theory).

The gauge-fixing fermion (\ref{(98)}) for our model is
\begin{equation}
\label{(186)}\Psi ^{\prime \prime }=-\int d^4x\left( \overline{\eta }%
_1^{ij}C_{ij}+\overline{\eta }_2^iA_{0i}+\lambda \frac{M^2}2\overline{\eta }%
_2^{ij}A_{ij}+\overline{\eta }_1^i\eta _{2_{0i}}+\frac 1\lambda u^i\overline{%
\eta }_i\right) ,
\end{equation}
where the bar variables belong to the non-minimal sector, while the ghosts $%
\eta _{2_{0i}}$ correspond to the gauge parameters $\epsilon _2^{0i}$. The
path integral in this case will read
\begin{equation}
\label{(187)}Z_{\Psi ^{\prime \prime }}=\int {\cal D}A^{ijk}\;{\cal D}\pi
_{ijk}\;\exp \left( i\overline{S}^{\prime }\right) ,
\end{equation}
where
\begin{equation}
\label{(188)}\overline{S}^{\prime }=\int d^4x\left( \dot A^{ijk}\pi
_{ijk}-3\pi _{ijk}\pi _{ijk}-\frac{M^2}{2\cdot 3!}A^{ijk}A_{ijk}+\frac
9{M^2}\left( \partial ^i\pi _{ijk}\right) ^2\right) .
\end{equation}
The gauge invariances of the total action in this case are obtained from the
extended ones making $\delta _\epsilon u^{ij}=\delta _\epsilon u^j=0$. They
take the form $\delta _\epsilon A^{\alpha \beta \gamma }=\partial
_{}^{[\alpha }\epsilon _2^{\beta \gamma ]}$, $\delta _\epsilon A^{\alpha
\beta }=\lambda \epsilon _2^{\alpha \beta }+\frac 12\partial _{}^{[\alpha
}\epsilon _2^{0\beta ]}$, $\delta _\epsilon \pi _{0ij}=\delta _\epsilon \pi
_{ijk}=\delta _\epsilon \Pi _{0i}=\delta _\epsilon \Pi _{ij}=0$, $\delta
_\epsilon v^{ij}=\ddot \epsilon _2^{ij}-\frac 12\partial ^{[i}\dot \epsilon
^{j]}$, $\delta _\epsilon v^i=\frac 12\left( \ddot \epsilon _2^{0i}+\lambda
\dot \epsilon ^j\right) $. In this case, the gauge transformations of the
fields $A^{\alpha \beta \gamma }$'s and $A^{\alpha \beta }$'s are manifestly
covariant too, due on one hand to the presence in (\ref{(179)}) of the
non-vanishing functions $C_{ij}^0$ and on the other to the constraints (\ref
{(182)}). The Lagrangian action of the reducible first-class family reads%
\begin{eqnarray}\label{(189)}
& &S_0^{\prime L}\left[ A^{\alpha \beta \gamma },A^{\alpha \beta }\right] =
\int d^4x\left( -\,\frac 1{2\cdot 4!}F_{\alpha \beta \gamma \rho }F^{\alpha
\beta \gamma \rho }\right) -\nonumber \\
& &\int d^4x\,\,\frac{M^2}{2\cdot 3!}\left( A^{\alpha \beta \gamma }-\frac 1
\lambda F^{\alpha \beta \gamma }\right) \left( A_{^{\alpha \beta \gamma }}-
\frac 1\lambda F_{\alpha \beta \gamma }\right) ,
\end{eqnarray} where $F^{\alpha \beta \gamma }=\partial ^{[\alpha }A^{\beta
\gamma ]}\equiv \partial ^\alpha A^{\beta \gamma }+\partial ^\gamma
A^{\alpha \beta }+\partial ^\beta A^{\gamma \alpha }$. Action (\ref{(189)})
allows the gauge invariances $\delta _\epsilon A^{\alpha \beta \gamma
}=\partial ^{[\alpha }\epsilon ^{\beta \gamma ]}$, $\delta _\epsilon
A^{\alpha \beta }=\lambda \epsilon ^{\alpha \beta }+\partial ^{[\alpha
}\epsilon ^{\beta ]}$. This action comes from the gauging of the rigid
symmetries $\delta _\epsilon A^{\alpha \beta }=\lambda \epsilon ^{\alpha
\beta }$ (here, $\epsilon ^{\alpha \beta }$ are all constant) of the action
\begin{equation}
\label{(190)}\tilde S_0\left[ A^{\alpha \beta }\right] =-\int d^4x\,\,\frac{%
M^2}{2\cdot 3!}\frac 1{\lambda ^2}F^{\alpha \beta \gamma }F_{\alpha \beta
\gamma }.
\end{equation}
We are under the conditions of Sec. 5.1, case ii) because $Z^\alpha \epsilon
^\beta =0$, for $\epsilon ^\beta $'s constant, with $Z^\alpha \equiv
\partial ^\alpha $. When $\epsilon ^\beta $'s depend on $x$ action (\ref
{(190)}) possesses some gauge symmetries independent of the presence of the
fields $A^{\alpha \beta \gamma }$, namely
\begin{equation}
\label{(191)}\delta _\epsilon A^{\alpha \beta }=\partial ^{[\alpha }\epsilon
^{\beta ]}.
\end{equation}
Formula (\ref{(191)}) is the analogous of (\ref{(146)}-\ref{(147)}) from the
general theory, in the case $\dot Z_{\;a_1}^a=0$. The conserved
gauge-invariant currents correspondent to the rigid symmetries $\delta
_\epsilon A^{\alpha \beta }=\lambda \epsilon ^{\alpha \beta }$ for action (%
\ref{(189)}) take the form
\begin{equation}
\label{(192)}j_{\,\,\,\,\alpha \beta }^\gamma =\frac{M^2}{3!}\left(
A_{^{\,\,\,\,\alpha \beta }}^\gamma -\frac 1\lambda ^{\,}F_{\,\,\,\,\alpha
\beta }^\gamma \right) .
\end{equation}
Action (\ref{(189)}) describes a field theory with abelian two and
three-form gauge fields coupled through a mixing-component term of the type
current-current, with the gauge-invariant current (\ref{(192)}).

The Wess-Zumino action in this case is precisely
\begin{equation}
\label{(193)}S_0^{\prime WZ}\left[ A^{\alpha \beta \gamma },A^{\alpha \beta
}\right] =-\frac{M^2}{12\lambda }\int d^4x\,F^{\alpha \beta \gamma }\left(
\frac 1\lambda F_{\alpha \beta \gamma }-2A_{^{\alpha \beta \gamma }}\right)
{}.
\end{equation}
More on abelian $p$-form gauge fields can be found in \cite{27}. This
completes our analysis.

\section{Conclusion}

In this paper it was shown in detail the way of quantizing the systems with
only second-class constraints in the BRST formalism by converting the
original second-class constraints into some first-class ones in a larger
phase-space. The main advantage of our method consists in the fact that it
is standard. Thus, the existence of the functions we are working with is
fully proved and, in addition, their concrete form is output. The way of
implementing the secondary first-class constraints exposed in this paper
emphasises the main difference between our conversion method and the BFT
method \cite{5}-\cite{6}. Indeed, the presence of the term $\gamma
_af^a\left( C\right) $ in the Hamiltonian of the first-class family is
decisive in underlining this difference as the functions $f^a\left( C\right)
$ are a characteristic of our method and do not appear in the BFT method. In
addition, we expose a conversion method for the reducible case, too.

At the same time, it is clarified the provenance of the first-class family
in the reducible, as well as irreducible case. As was exhibited, the
first-class family results from the gauging of some rigid symmetries of a
certain action. In the context of building up this family, the Wess-Zumino
action appears natuarally, its concrete form being computed in both cases.

The two examples illustrating the theoretical part of the paper also
evidence that our method lead to a manifestly covariant form of the
Lagrangian action corresponding to the first-class family. The presence of
the non-identically vanishing functions $C_a^0$ in (\ref{(24)}), and
implicitly in (\ref{(35)}) is crucial in order to establish the covariance.

\end{document}